%% file: cmtydef.tex
\pdfoutput=1

\documentclass[twocolumn,prX,english,nofootinbib,superscriptaddress]{revtex4}
\newcommand{\halfwidth}{4cm} \newcommand{\wholewidth}{7cm} \newcommand{\threehighwidth}{6cm}

\usepackage{docmute}

\usepackage[utf8]{inputenc}
\usepackage{grffile}
\usepackage{graphicx}
\usepackage{subfigure}
\usepackage{amsfonts}

\usepackage{calc}
\usepackage{color}

\usepackage[normalem]{ulem}

\begin{document}

\title{An edge density definition of overlapping and weighted graph communities}
\author{Richard K. Darst}
\author{David R. Reichman}
\affiliation{Department of Chemistry, Columbia University, 3000
  Broadway, New York, NY 10027, USA.}
\author{Peter Ronhovde}
\author{Zohar Nussinov}
\affiliation{Department of Physics, Washington University in St. Louis,
Campus Box 1105, 1 Brookings Drive, St. Louis, MO 63130, USA}
\begin{abstract}
  Community detection in networks refers to the process of seeking
  strongly internally connected groups of nodes which are weakly externally
  connected.
  In this work, we introduce and study a community definition based on internal edge
  density.  Beginning with the simple concept that edge density
  equals number of edges divided by maximal number of edges, we apply
  this definition to a variety of node and community arrangements to
  show that our definition yields sensible results.  Our
  community definition is equivalent to that of the Absolute Potts
  Model community detection method (Phys. Rev. E 81, 046114 (2010)),
  and the performance of that method validates the usefulness of
  our definition across a wide variety of network types.
  We discuss how this definition can be extended to weighted,
  and multigraphs, and how the definition is capable of
  handling overlapping communities and local algorithms.
  We further validate our definition against the recently
  proposed Affiliation Graph Model (arXiv:1205.6228 [cs.SI]) and show
  that we can precisely solve these benchmarks.
  More than proposing an end-all community definition, we explain how
  studying the detailed properties of community definitions is
  important in order to validate that definitions do not have
  negative analytic properties.  We urge that community definitions be
  separated from community detection algorithms and and propose that
  community definitions be further evaluated by criteria such as
  these.
\end{abstract}
\maketitle

\section{Introduction}

It is standard to use the concepts of graph theory in order to
represent the interactions of complex systems.
Here, the nomenclature of ``nodes'' and ``edges'' is used to
represent generic items and the interactions between them\cite{bollobas1998modern}.
One important form of structure in graph theory is that of
\emph{communities}, or strongly connected subgroups\cite{fortunato2010community}.  There is no
single agreed upon definition for communities, but it is generally
accepted that communities are
groupings of nodes that are strongly connected to each other
and weakly connected
to nodes in other communities.  It is important
to note the two uses of the term ``community'' here.  The first is a
real-world grouping of objects, sometimes known as ``ground-truth'' in
other works\cite{yang2012defining}.  This grouping is not precisely mathematically
defined, but is empirically defined based on the, e.g., social,
anthropological, biological, etc. data upon which the graph is
created\cite{mislove2007measurement,
zachary1977information,lusseau2003emergent,
palla2005uncovering,prill2005dynamic,
i2001topology,calvert1997modeling}.
The second form of ``community'' is a
mathematical construction of nodes and edges.  The community
detection field has two goals: the definition of mathematical communities
that most correspond to real-world communities, and the
development of algorithms that can locate these mathematical
communities within graphs.

Many community definitions have been proposed in the literature.  Several
reviews are dedicated to an
overview and comparison of community detection (CD) methods, and more
specifically, of community
definitions themselves  \cite{fortunato2010community,
  yang2012defining,radicchi2004defining}.
Often, primary emphasis is placed on the
description and workings of the community detection method itself, and
the method's particular community definition is only
implicitly defined as the practical result of applying the CD algorithm, and
not as a separate formal definition.  This has been accepted as a
practical thing to do, but much more could be learned if the scope and definitions of communities were broader.

The Girvan-Newman modularity is one of the few heavily-studied community
definitions \cite{girvan2002community,
  newman2004finding, newman2004fast, lehmann2007deterministic,
  fan2007accuracy}.  Modularity weights internal edges
against external edges in an attempt to indicate, without any user-input
parameters, a ``best'' community structure for any type of graph.
Modularity is one of the oldest and most-emphasized of the existing
community definitions, but
despite its utility, there is a useful caveat to be made in the study of
(ground-state) definitions of modularity.
In 2006 Fortunato and Barth\'elemy showed that modularity
has a very interesting property: namely an implicit dependence on the
total size of the network to which it is applied\cite{fortunato2007resolution,lancichinetti2011limits}.  This prevents
modularity-based community detection methods from
resolving small communities in a large
graph.  In short, the optimal
community in one partition of a graph depends upon properties of the graph
far away.  This behavior is non-intuitive and referred to as a
``resolution limit.''  Several attempts have been made to produce
multi-resolution modularity measures which have a tunable parameter
that ``zooms'' in or out and controls the size of detected communities
\cite{reichardt2006statistical}, but these too have been shown to
suffer resolution limits
\cite{kumpula2007limited, guimera2004modularity}.  Modularity has taught us several things.
First and foremost,
community definitions need theoretical study.  Second,
\emph{any} global community definition may have a resolution
limit, necessitating \emph{local} community definitions \cite{kumpula2007limited,reichardt2004detecting,good2010performance,lancichinetti2011limits}.

In order to understand the relevance of the work presented in this
work, it is useful to look at past community detection approaches from
the standpoint of optimizing over some configurational space.  Instead
of trying to improve sample techniques, we are looking at a different
cost function which gives a \emph{different ground state}, which may
be ``more correct'' than another ground state.  As an example, after
the introduction of modularity as a measure of community structure,
many attempts were put forth to improve community detection via the
use of enhanced sampling of the configuration space of community
assignments in order to better optimize modularity
\cite{duch2005community, hofman2008bayesian, lehmann2007deterministic,
  schuetz2008efficient, newman2004fast, brandes2008modularity,
  blondel2008fast, pujol2006clustering, liu2010advanced,
  noack2009multi, agarwal2008modularity}.  As useful as these methods
are, they all share one fundamental limit: they rely on the assumption
that modularity is the correct cost function to optimize and that
community detection is limited by the ability to properly sample
configuration space \cite{guimera2004modularity}: basically,
overcoming kinetic barriers in minimization dynamics.  On the other
hand, it is natural to place an emphasis on understanding other
\emph{ground state} community definitions and properties, i.e. the
mapping from real-world to mathematical communities, before focusing
on the search for optimal partitions.  The viewpoint we espouse is
that only after understanding ground state characteristics should one
focus on the sampling required to finding that ground state.  There
are few such analyzes in existing literature, but the number is
growing.

In this work, we propose a community definition based on \emph{internal
  edge density} (abbreviated as simply ``edge density'').
Edge density-based definitions have been considered before in a
wide variety of contexts
\cite{yang2012defining,radicchi2004defining,
mancoridis1998using,wasserman1994social,scott2000social,moody2003structural,
luce1949method}.
The particular edge density
definition we focus on has been used in an implicit fashion previously
\cite{ronhovde2009multiresolution, ronhovde2010local},
but has never been explicitly defined and extensively studied, as we do
here.  Our definition affords us the ability to \emph{compare} the
properties of graph-based communities to the properties of real-world
communities as a means of quantifying the degree to which they match.
Currently the most common method of assessing
community detection algorithms is to choose a test graph and run
community detection.  This gives a biased perspective of community
detection, since it takes some initial belief about the structure of
communities and optimizes methods towards that definition.  Recently,
there has been much needed emphasis on the comparison of community
definitions to real-world communities.  In order to make progress, we
show how edge
density relates \emph{to actual} properties of model graphs and communities.

In addition to the above benefit of edge density-based communities,
edge density can be quite naturally
extended to weighted and multi-graphs, and handle
overlapping communities, all important areas of modern research in
community detection.
In particular, few algorithms have been proposed that are capable of
detecting overlapping communities.  Thus, our work not only provides a
conceptual breakthrough in terms of a rigorous analysis of new
community definitions, but also provides a practical benefit in community
detection ability.  As an outgrowth of these extensions, we propose a new
\emph{variable topology Potts model}, which allows a more natural means of
community detection in \emph{heavily weighted graphs}.

To provide concrete illustration of these claims, we turn to a recently
proposed proposed social network model.  In particular, we focus on
recent work of
Yang and Leskovec (YL), who have proposed a new model of social
networks, the \emph{Affiliation Graph Model} (AGM)\cite{yang2012structure}.  This model takes
into account features observed in real-world social networks found via
comparison with online social networks.  YL claim that no
current community detection algorithms can describe and detect
communities in these graphs.  We conceptually show that the edge
density community
definition models this graph properly, and then perform actual
community detections to prove that this is the case.

We do not claim here that edge density is a universal definition of community that
applies to all possible communities in all possible graphs.
Instead, we provide the tools to determine if any one particular class
of graphs is well described by an edge density picture, either by an
analysis of the graph generation process, or properties of the edge
structure.  We hope to inspire similarly detailed analysis, and more
rigorous comparison, of existing and future community definitions.

This remainder of this work is organized as follows: In
section~\ref{sec:ED-edgedensity}, we define the basic tools needed to
quantify edge density, then state our edge density definition. In
section~\ref{sec:ED-edgedensitymodels}, we describe historic and new
edge density models, and the Potts model framework we use to perform
actual community detection.  In section~\ref{sec:ED-modelindependent},
we describe some universal properties of edge density, which are
independent of the exact model used to perform community detection.
In section~\ref{sec:ED-modeldependent}, we describe how certain models
handle the boundaries between communities differently, and explain our
model of choice.  In section~\ref{sec:ED-algorithms} and
Appendix~\ref{sec:ED-app-algorithms}, we discuss practical
considerations for constructing an actual algorithm employing edge
density.  In section~\ref{sec:ED-edgedensityscaling}, we validate the
use of edge density as a scale parameter.  In
section~\ref{sec:ED-overlaps}, we discuss general considerations for
overlapping communities, and contrasting with some recent work, we
show that edge densities detect the correct conceptual behavior.  In
section~\ref{sec:ED-AGM}, we rigorously correlate our edge density to
affiliation graph models, and see that we can detect these communities
easily.  In section~\ref{sec:ED-weighted}, we describe the extension
to weighted graphs, and how our new variable topology Potts model
provides a significant improvement over older models.  In
section~\ref{sec:ED-directedetc}, we discuss various possible
extensions. In section~\ref{sec:ED-discussion}, we discuss some
general background and address some possible limitations of edge
density.  In section~\ref{sec:ED-overoptimization}, we discuss the
danger of over-optimizing to particular benchmark models and why
studies of edge density are useful nonetheless.  In
section~\ref{sec:ED-conclusions}, we provide concluding remarks and
discuss directions for future research.

\section{Edge density}
\label{sec:ED-edgedensity}

We must first define the nomenclature and measures we will use.  We
will specify communities or groups of nodes by capital latin letters
such as $A$ and $B$, and individual nodes by lowercase letters such as
$a$, $b$, $x$, and $y$.  In particular, the node $a$ shall represent an
arbitrary node of community $A$, and specific nodes of community $A$
are indexed as $a_i$.  The nodes $x$ and $y$ generally represent nodes which
are not currently in any community.  The community $A$ will have $n_A$
nodes.

We use a set terminology to discuss communities and groups of
nodes.  Set union
is denoted with ``$\cup$'', indicating every node on the left or
right or both sides of the operator.  Set intersection is denoted
``$\cap$'', indicating
every node that is on both sides of this operator.  We use this
nomenclature loosely, allowing constructs such as ``$A \cup x$'' even
though we are operating on a community on the left and an individual
node on the right.  This should be taken to mean union of the set of
nodes in community $A$ and the set of nodes containing $x$.

For any group of nodes (explained below), $l$
represents the maximal number of possible edges (``links'' in
our nomenclature).  The variable $e$ represents the actual number
of edges.  The \emph{edge density} $\rho$ is the fraction of these
links which have edges actually present,
\begin{equation}
  \label{eq:edgedensity}
  \rho = \frac{e}{l}.
\end{equation}

\begin{figure}
  \includegraphics[width=\wholewidth]{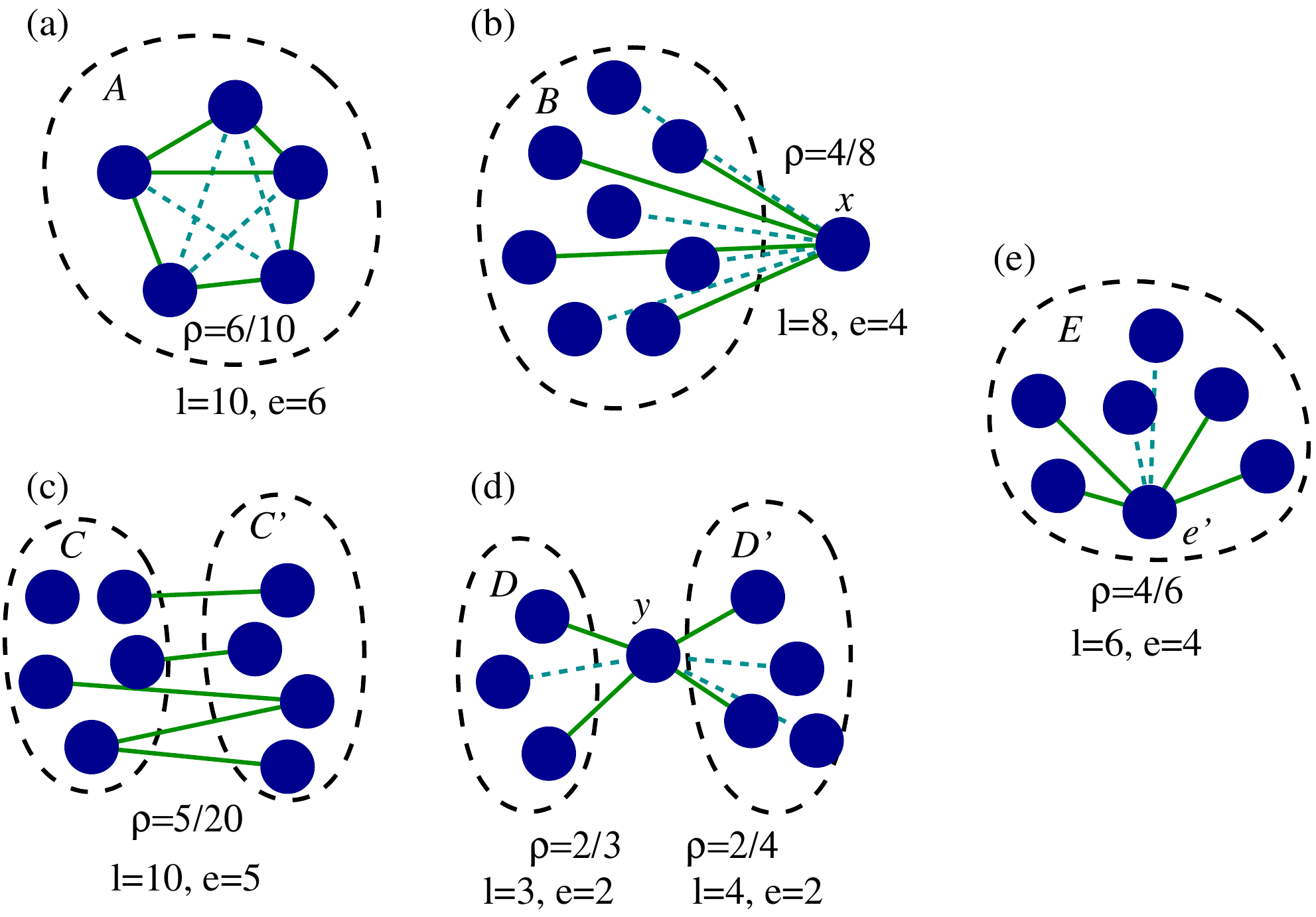}
  \centering
  \caption[Edge density examples]{Edge density calculations in a
    variety of situations.  For each situation, $l$ is the maximal
    possible number of edges, $e$ is the actual number of edges, and
    the edge density $\rho=e/l$.
    In (a), we see the edge density of one community $A$, a
    group of $n=5$ nodes, with $l = \frac{1}{2}n(n-1)=10$ total
    possible links (dotted and solid lines), $e=6$ actual edges (solid
    lines), thus leading to a $\rho_A=\frac{6}{10}$.  In (b), we see
    the edge density definition applied between one node $x$ and a
    community $B$.  The community $B$ contains eight nodes, for a
    total number of possible links to node $x$ of $l_{Bx}=8$.  We see
    four actual links in place, leading to a $\rho_{Bx}=\frac{4}{8}$.
    Note that the internal connectivity of the community $B$ is
    irrelevant and not shown.  In (c), we see the edge density as
    applied to two communities $C$ (5 nodes) and $C'$ (4 nodes).
    Internal edge structure is again irrelevant and not shown.  There
    are a total of $l_{CC'}=n_C n_{C'}=20$ possible links between
    these two communities, $e_{CC'}=5$ edges in place, for an edge
    density of $p_{CC'}=\frac{5}{20}$.  (d) shows the edge density of
    one node $d$ being pulled between two communities $D$ and $D'$.
    Edge density for each half is in analogy to (b).  In (e), we see
    the edge density between a node $e$ and a community which contains
    it, $E$.  This is similar to (b) except that, by convention, we
    use $n_E-1$ links as our basis.  \label{fig:vardefs}}
\end{figure}

We can calculate the edge density within and between a variety of
different types of groupings of nodes, all of which can be useful
under different circumstances.  Fig.~\ref{fig:vardefs} illustrates the
most common situations.
(a) We can calculate the edge density within only one
community $A$, in which case $l=\frac{1}{2}n_A(n_A-1)$.  Community
edge density is used
to quantify the absolute community size and scale.
(b) We can calculate the edge density between a community $B$
and a node $x$ not currently in community $B$, in which case
$l=n_B$, since that node can connect once to every node in $B$.
This is useful when deciding if a node should enter or leave a
community.
(c) We can calculate
the edge density between two communities $C$ and $C'$, in which case
$l=n_C n_{C'}$, since every node in $C$ can connect once to every node in
$C'$.  This is useful when testing if two communities should
merge.
(d) We can calculate the edge density for the case of one
node $y$ which has edges between two communities $D$ and $D'$, in which
case $l_{By}=n_B$ and $l_{D'y}=n_{D'}$.  This is
useful for deciding which of two communities the node $y$ would prefer
to join.
(e) By convention, when we calculate the edge density
between a community $E$ and node $e'$ which is currently within that
community, we only consider
links between $e'$ and the $n_E-1$ other nodes of the communities.
Thus, in this case, $l_{Ee'}=n_E-1$ instead of $n_E$, in contrast to
case (b).

\subsection{The edge density community definition}
\label{sec:EDCD}

The most basic form of an edge density community definition (EDCD) for
overlapping or isolated communities consists
of three parts.  (a) A community $A$ of scale $\rho^*$ is \emph{a subset
  of nodes with an internal edge density $\rho_A$ greater than $\rho^*$}.  (b)
Communities are taken to be as large as possible (absorbing as many nodes as
possible) as long as
$\rho > \rho^*$.  (c) A community $A$ of scale $\rho^*$ must have each
individual node $a$'s edge density $\rho_{Aa} > \rho^*$.

Part (a) is the essence of the definition, establishing a scale parameter of
the communities.  Part (b) is necessary because without it, we could
always prefer smaller, more dense communities.  For example, a graph
could be partitioned into two-node cliques, all of which have an edge
between the nodes, to get communities which all have $\rho=1$ but do
not provide useful information concerning the overall structure of the
graph.  Part (c) is
necessary to ensure that every node is well-connected to the
community, preventing nodes only loosely connected from being added to
any community.
In future sections, we will rigorously show that the absolute Potts
model and the variable topology Potts model return communities
satisfying these criteria.  Part (a) will be shown via
Eqs.~(\ref{eq:apmPStar}, \ref{eq:VTPMpstar}) in
Sec.~\ref{sec:ED-edgedensitymodels}.  Parts (b) and (c) are discussed
in Sec.~\ref{sec:pottsEnergyPerturbation}.

This definition is sufficient when
communities are considered in isolation, i.e. when we never consider one
node and must decide which of several communities it might join, if it
would increase $\rho$ for any of them (Fig.~\ref{fig:vardefs} (d)).
In these cases, one may choose to add the node to a larger community
resulting in a still larger resultant community, or to the community to
which it shares the higher edge density.  Both of these are reasonable
options.  One simple criteria would be to use the total number of
edges as a criteria: a node $x$ joins community $A$ instead of
community $B$ if $e_{Ax} > e_{Bx}$.  Existing edge density community
definition methods make different choices here, which will be
discussed in Section~\ref{sec:ED-modeldependent}.

\subsection{Connection to graph generation processes}
\label{sec:ED-binomial}

The edge density variable $\rho$ refers to actual edge densities in a
specific graph instance.  Many benchmark graphs are designed
stochastically, with edges placed within or between communities with a
some specified probability of $p$.  Since each link (possible edge)
has an edge placed with a probability of $p$ independent of all other
edges, the variable $e$ (number of edges placed) is
binomially distributed with mean $lp$ (and probability of success
$p$).  Each of the situations depicted in Fig.~\ref{fig:vardefs} can
have a defined $p$.  For example, \emph{planted l-partition model}
(also known as \emph{Stochastic Block Models} (SBMs))
graphs are defined by specifying intra-community edge densities
$p_A$, $p_B$, etc. (Fig.~\ref{fig:vardefs} (a)), and inter-community
edge densities $p_{AB}$, $p_{AC}$, $p_{BC}$,
etc. (Fig.~\ref{fig:vardefs} (c)) \cite{condon2001algorithms}.
Then, each respective $e$ will
have a binomial distribution with (its respective) $l$ trials and (its
respective) $p$ chance (of an edge) per trial,
\begin{equation}
  P[e=x] = \mathcal{B}(x; l, p) \approx \mathcal{N}(x; lp, lp(1-p)).
\end{equation}
$\mathcal{B}(l,p)$ represents a
binomial distribution of $l$ trials and $p$ probability per trial, and
$\mathcal{N}(
\langle x \rangle, \sigma^2)$ represents a normal distribution with a
mean of $\langle x \rangle$ and a variance of $\sigma^2$ after we make a normal
approximation to the binomial distribution.
With $\rho=e/l$, we have a distribution of $\rho$ of
\begin{eqnarray}
  P\left[\rho=\frac{x}{l}\right] &=& \mathcal{B}(x; l, p), \\
  P\left[\rho=x\right] &\approx& \mathcal{N}\left(x ; p, \frac{p(1-p)}{l}\right).
\end{eqnarray}

In the above, we employed the following shorthand for (a normalized) binomial distribution,
\begin{eqnarray}
{\mathcal{B}(m;n,p)} = { {n} \choose {m}} p^{m} (1-p)^{n-m}.
\end{eqnarray}
The normal distribution is, explicitly, given by
\begin{eqnarray}
\mathcal{N}(x ; \left< x \right>, \sigma^{2}) = \frac{1}{\sqrt{2 \pi \sigma^{2}}} \exp\left[- \frac{[x - \left< x \right>]^{2}}{(2 \sigma^{2})}\right].
\end{eqnarray}

Clearly, $\rho$ is a random variable with a mean of $p$ and has a
standard deviation of $\sqrt{\frac{p(1-p)}{l}}$.  We see the intuitive
result that as we approach larger community sizes, we approach
mean-field $\rho = p$ with decreasing standard deviation around this
mean.  This relation between $\rho$ and an underlying $p$ is valid for
any case where all links share the same edge probability.

\section{Edge Density Models}
\label{sec:ED-edgedensitymodels}

In the previous section, we stated a basic edge density definition.
This definition has been implicitly used in several preexisting
community detection methods.  Thus,
the edge density community definition is not new, but it has never
been as explicitly stated and analyzed.  Before we proceed to more
specific uses of
the edge density community definition, we will review existing
Potts-model based edge density community definitions.

Recent edge density-based community definitions are formulated in
terms of \emph{Potts models} \cite{wu1982potts,potts1952generalized,ashkin1943statistics}.  The Potts model is a
spin system, with
each site having one of up to $q$ associated spin flavors $\sigma = 0, \ldots, q-1$.
It is similar to the Ising model except that the Ising model allows
spins of only $0$ or $1$ (i.e., the Ising model is a Potts model with $q=2$).  For community detection, we take the spin flavor
of each site (node) as corresponding to a community assignment. That is, if $\sigma_{a} = p$ then node $a$ lies in community number $p$.
(The index $p$ clearly lies in the range $0\le p \le (q-1)$). Inter-site interactions which define a Hamiltonian are
minimized to find an optimal community assignment.  This may be done
by penalizing inter-community edges and favoring intra-community
edges.  Reichardt and Bornholdt explain the balances and equivalences
of sums of internal and external edges in
\cite{reichardt2006statistical}.  The Reichardt-Bornholdt (RB) Potts model was the first explicit Potts
model used for community detection, but it does \emph{not} use our
edge density community definition \cite{reichardt2004detecting,
  reichardt2006statistical}.  Instead, the RB Potts model is
equivalent to the Girvan-Newman modularity.  Because the RB Potts
model does not correspond to the edge density definitions in this
paper, it is not considered in the following analysis.

Potts models are ideal for edge density definitions, because they consist
primarily of a sum over all edges \cite{heimo2008detecting}.  The models allow us to select only
internal edges, although it is equally possible to select only external
edges.  One possible limitation of Potts models is that each node can
only be associated with one community.  This would seem to exclude the
possibility of overlapping communities.  In order to expand the Potts
models to allow overlapping communities, we reformulate the models
away from an
edge-centric definition towards an equivalent community-centric
definition.

\subsection{Absolute Potts model}

The absolute Potts model (APM) is a spin-based ferromagnetic community
detection method\cite{ronhovde2009multiresolution, ronhovde2010local}.
It has been shown
to be very accurate at community detection under a wide variety
of circumstances\cite{lancichinetti2009community,
  ronhovde2009multiresolution, ronhovde2010local, ronhovde2011detecting}.  The APM implicitly uses the edge density community
definition.  In the Potts representation of
the APM, the CD problem is mapped onto an energy minimization of a
Hamiltonian defined over pairs of spins
\begin{equation}
  \label{eq:APMspin}
  E = - \sum_{a, a'\neq a} \frac{1}{2}
    \left(A_{aa'} - \gamma B_{aa'} \right)
      \delta(\sigma_a, \sigma_{a'}).
\end{equation}
The sum above is over distinct nodes $a$, $a'$ with, as explained above, $\sigma_a$ denoting the
community assignment of node $a$, and $A_{aa'}$ being the weighted
``adjacency matrix''.  In unweighted graphs, $A_{aa'}=1$ if an edge is present between nodes $a$
and $a'$, and $A_{aa'}=0$ otherwise.  For weighted graphs, described further
in Sec.~\ref{sec:ED-weighted}, $A_{aa'}=w$, where $w$ is
the respective edge weight. In unweighted graphs, we will define 
the ``inverse adjacency matrix'' to have elements of $B_{aa'}=1$ if there is
no edge preset, zero otherwise.  For more general unweighted graphs, we will set
$B_{aa'}=1-A_{aa'}$.
In these models, lower energies are attractive, and any shifts in
community assignment that lower energy are favorable.

This Hamiltonian in Eq. (\ref{eq:APMspin}) involves a sum over all
edges. However, due the presence of the Kronecker delta
(i.e., $\delta(\sigma_a \sigma_{a'})=1$ if
$a=a'$ and zero otherwise), the sum is inherently local.
That is, the sum contains
only edges between nodes which are in the same
community.  An inter-community edge has a favorable (negative) energy
of $1$, and each missing edge has an energy penalty (positive energy)
of $\gamma$.
While typically represented within a spin formulation, this can be recast
with the primary sum over communities instead of over edges:
\begin{equation}
  \label{eq:APMcmty}
  E = - \sum_{A} \sum_{(a,a'\neq a) \in A}
      \frac{1}{2}\left( A_{aa'}-\gamma B_{aa'} \right),
\end{equation}
with $(a, a')$ being all ordered pairs of nodes in community $A$.  By
the use of the term ``ordered'', we make evident that we include both the pair $(a,a')$ and the pair $(a',a)$ (for $a\neq a'$) in the sum.
The spin formulation appears simpler
conceptually, and may be implemented much more efficiently, although
the community-centric definition allows us to make additional theoretical
progress.

Since the inner sum of $\frac{1}{2}A_{aa'}$ is the number of
edges in community $A$, and the sum of $\frac{1}{2}B_{aa'}$ is the
number of missing edges in community $A$, we can rewrite this as
\begin{equation}
  E = - \sum_{A} \left( e_A - \gamma(l_A-e_A) \right),
  \label{eq:APMcmty2}
\end{equation}
where the inner sum has been rewritten in terms of the number of
existing edges $e_A$ and the number of missing edges $l_A-e_A$.  With
minor rearrangement and invoking our edge density definition $\rho=e/l$, we
may rewrite the energy in terms of edge density with a prefactor of $l_A$,
\begin{equation}
  \label{eq:apmPgamma}
  E = - \sum_{A} l_A \left( \rho_A - \gamma(1-\rho_A) \right).
\end{equation}
The energy is now written as \emph{sum over edge densities}.  We see that for the
energy of any one community to be negative (attractive and having a
binding energy), the term $\rho_A -
\gamma(1-\rho_A)$ must be positive.  Rearranging, this gives us a
correspondence between the APM variable $\gamma$ and the critical
(minimum) edge density $\rho^*$
\begin{eqnarray}
  \rho_A &>& \frac{\gamma}{1+\gamma} = \rho^* \label{eq:apmPStar}.
\end{eqnarray}
Eq.~(\ref{eq:apmPStar}) is the fundamental relationship between the APM variable
$\gamma$ and the edge density critical value $\rho^*$.  It shows that
any community returned by the APM must satisfy the edge density
community definition part (a).
We can now
state two identical relationships which give equivalences of the APM and the
edge density viewpoints:
\begin{eqnarray}
  \rho^* &=& \frac{\gamma}{1+ \gamma} \label{eq:gammaToRho}, \\
  \gamma &=& \frac{\rho^*}{1-\rho^*}  \label{eq:rhoToGamma}.
\end{eqnarray}

We can rewrite the APM Hamiltonian in terms of $\rho^*$ instead of
$\gamma$ using Eqs.~(\ref{eq:apmPgamma}, \ref{eq:rhoToGamma}),
\begin{equation}
  E = - \sum_{A} l_A \left( \frac{\rho_A - \rho^*}{1-\rho^*} \right).
  \label{eq:APMHamiltonian}
\end{equation}

We can now relate the previously existing APM to our new edge density
community definition.  First, if $\rho_A>\rho^*$, the energy
of community $A$ is negative, and thus has a binding energy.
Therefore, all communities $A$ must have $\rho_A > \rho^*$, corresponding to
the edge density community definition part (a).  Second, $1/(1-\rho^*)$
and $l_A=\frac{1}{2}n_A(n_A-1)$ are scale factors.
In order to minimize energy (and create
the best partition according to the APM), we want $l_A$ to be as large
as possible (larger communities), and also $\rho_A$ to be as large as
possible.  Furthermore, because of the factor $(\rho_A-\rho^*)$, increasing
the edge density of the community also results in lower energy.
As we will see in Sec.~\ref{sec:ED-edgedensityscaling}, larger community sizes generally
tend to imply smaller $\rho$, thus larger community size and larger
$\rho$ are competing factors which must be balanced.
Eq.~(\ref{eq:APMHamiltonian}) quantifies that balance.
Communities with single nodes (``size one'' communities) have $E=0$
according to Eq.~(\ref{eq:APMspin}), and this can be represented in
Eq.~(\ref{eq:APMHamiltonian}) if we define a community
consisting of only one node to have $\rho = 1$.

\subsection{Variable topology Potts model}

The absolute Potts model
\cite{ronhovde2009multiresolution,ronhovde2010local}
has, by now, been introduced and studied in earlier works.  The variable topology Potts model (VTPM), 
which we now introduce, has been
hinted at in previous literature,
but never defined nor extensively analyzed\cite{traag2011narrow, reichardt2004detecting}.
A somewhat similar technique was used in
Ref.~\cite{ronhovde2011detecting}, where \emph{weights} were shifted by
a value $\overline{V}$.  The approach of Ref.~\cite{ronhovde2011detecting}
technique requires an adjustment of two variables, $\overline{V}$ and
$\gamma$, in order to find optimal partitions, while the technique of
this section only requires adjustment of $\rho^*$.

The defining property of the variable topology Potts model is that it assigns a constant penalty (of
size $\rho^*$) to all links, as opposed to only missing edges.
The VTPM Hamiltonian is
\begin{equation}
  \label{eq:VTPMspin}
  E = - \sum_{a, a'\neq a} \frac{1}{2}
    \left(A_{aa'} - \rho^* \right)\delta(\sigma_a, \sigma_{a'}).
\end{equation}
As in the APM, this can be recast into a form which sums over
communities first, instead of over all pairs of nodes
\begin{equation}
  E = - \sum_{A} \sum_{(a,a'\neq a) \in A}
      \frac{1}{2}\left( A_{aa'}-\rho^* \right),
\end{equation}
with all variables analogous to the APM nomenclature in
Eq.~(\ref{eq:APMcmty}).
We proceed as in the APM to turn this into an edge-density based
definition using the same relation $e_A = \sum A_{aa'}$
\begin{eqnarray}
  E &=& - \sum_{A} l_A
      \left( \rho_A - \rho^* \right). \label{eq:VTPMHamiltonian}
\end{eqnarray}
We, again, have a criteria which must be satisfied for any VTPM
community to have a binding energy,
\begin{equation}
  \rho_A > \rho^*    \label{eq:VTPMpstar}
\end{equation}
which is identical to the criteria of the APM and of edge density
community definition (a).
In this form, we see that the VTPM insists that all communities have
an edge density greater than $\rho^*$, which very naturally corresponds to
the edge density community definition.  Communities are weighted
by the number of links $l_A = \frac{1}{2} n_A (n_A-1)$.

It is illuminating to contrast the VTPM Hamiltonian of
Eq.~(\ref{eq:VTPMHamiltonian}) with the
APM Hamiltonian of Eq.~(\ref{eq:APMHamiltonian}).  They appear identical,
with the exception of the APM's inclusion of a scale factor of
$\frac{1}{1-\rho^*}$.  This means that \emph{all past work on the APM
  applies equally to the VTPM}.  In particular, we do not need extensive
additional testing on the VTPM in order to show that it achieves the
same performance.
The APM scale factor of $\frac{1}{1-\rho^*}$ becomes negative
when $\rho^*>1$.  In all of our analysis thus far, this would not
seem to be a limitation, but when we consider weighted graphs in
Sec.~\ref{sec:ED-weighted}, it will.  This is the first advantage of the
VTPM over the APM.

A ``clique'' is a group of nodes all mutually connected (thus,
$\rho=1$).  Normally, community detection methods do not break up
cliques, because they are as strongly connected as possible.  However,
if the edges are weighted, there may be certain ``weak'' edges where
it is reasonable to separate communities.  In the APM, weighted
cliques can not be subdivided, since it only places energy penalties
at \emph{missing} edges.  The VTPM overcomes this limitation by
allowing the least weighted edges to become repulsive repulsive first
as $\rho^*$ is increased.  Cliques can then be subdivided at their
weakest point.  The VTPM is so named because it is able to ``change
the topology'' of these cliques.

The core advantages of the VTPM over the APM are the ability to break up
cliques and handle weighted graphs more naturally.  Otherwise, it contains the same
information as the APM.  Instead of the control parameter
$\gamma = \frac{\rho^*}{1-\rho^*}$, it uses $\rho^*$ directly, leading
to a much more natural interpretation of the community scale.  Because
of this, the VTPM provides a compelling community detection algorithm
for future use, which will be elaborated on in future sections.

\subsection{Discussion of Potts models}

The APM has been extensively studied and shown to have acceptable
performance in community detection
across a wide variety of conditions and problems,
including common equal-sized and power law distributed graphs
\cite{lancichinetti2009community,ronhovde2010local}.  Since
we have shown that the APM directly uses the edge density community
definition (although unstated
explicitly until now), even without additional tests, we have strong
support for the edge density community definition.
While the VTPM has not been as extensively studied, by comparing
Eq.~(\ref{eq:APMHamiltonian}) and
Eq.~(\ref{eq:VTPMHamiltonian}), we see that the Hamiltonians are the
same for unweighted graphs, save a
scale factor of $\frac{1}{1-\rho^*}$.  This means that the VTPM also
will be equally powerful in all of the above cases.

\subsection{Energy changes upon community assignment perturbation}
\label{sec:pottsEnergyPerturbation}

The APM and VTPM are sums over all edges provided they connect nodes
in the \emph{same community}.  Thus, when we make some change to the
community assignments, for example, by combining two communities, the
energy change is completely represented by the sum of energies from
edges which were just moved into the same community, minus sum of energies
from edges which have been removed from the same community.  As an
example of this, consider adding node $x$ to community $B$, Fig.~\ref{fig:vardefs} (b).  The
entire APM energy remains unchanged except for the $n_B$ links between $x$
and the nodes of $B$.  Thus, the energy change can be represented as
\begin{eqnarray}
  \Delta E &=& \sum_{b \in B} \left( A_{bx} - \gamma B_{bx} \right)\nonumber \\
           &=& e_{Bx} - \gamma (l_{Bx} - e_{Bx} ) \nonumber \\
           &=& l_{Bx} \frac{\rho_{Bx} - \rho^*}{1-\rho^*}.
                                                  \label{eq:edgeperturbation}
\end{eqnarray}
In the above, $A_{bx}$ and $B_{bx}$ correspond to the adjacency and
inverse adjacency matrices as defined previously, while other uses
of $B$ within subscripts correspond to the community $B$ in
Fig.~\ref{fig:vardefs}.
For the VTPM, the energy change upon perturbation is also analogous,
\begin{equation}
  \label{eq:edgeperturbation-VTPM}
  \Delta E = l_{Bx} \left( \rho_{Bx} - \rho^* \right).
\end{equation}
The energy changes via perturbation have the same form as for the global
energy, with the sum only over the subset of edges created, and
the subtraction of only
a sum over the subset of edges removed.  The local nature of
energy makes it fast to compute energy changes under system
perturbations.  Community detection then
becomes a fairly well understood problem of sampling an energy
landscape with local interactions, using dynamics of the user's
choice.

Using Eq.~(\ref{eq:edgeperturbation}), we can see that a node $x$ will
join a community $B$
(Fig.~\ref{fig:vardefs} (b)) if the node-to-community edge density is
$\rho_{Bx}>\rho^*$.  A node $e'$ will \emph{leave} a community $E$
(Fig.~\ref{fig:vardefs} (e)) if the node-to-community edge density is
$\rho_{Ee'}<\rho^*$.  We can also show that two communities $C,C'$
will merge (Fig.~\ref{fig:vardefs} (c)) if the inter-community edge
density is $\rho_{CC'}>\rho^*$.  Thus, edge density naturally describes all
possible community merges, with $\rho^*$ being the critical edge
density for all possible merges or splits.

The above shows that communities returned by the Potts models satisfy
(b) and (c) of the edge density community definition,
Sec.~\ref{sec:EDCD}.
Criteria (c) states that community $B$ will have every node $\rho_{Bx}
> \rho^*$.  If this was not true in a community, energy would be
lowered by removing node $x$ from the community via the inverse of
Eqs.~(\ref{eq:edgeperturbation},\ref{eq:edgeperturbation-VTPM}).
Criteria (b) states that a community $B$ will grow as long as $\rho_B$
stays greater than $\rho^*$.  This is evidenced by the fact that any
node $x$ with
$\rho_{Bx}>\rho^*$ will have a favorable energy change upon joining
$B$, growing $B$ is large as possible as long as all nodes have
sufficient node-to-community edge density.
Eq.~\ref{eq:rho-community-mean} ensures that the resulting community
has $\rho_B>\rho^*$.

\section{Model-independent community properties}
\label{sec:ED-modelindependent}

The edge density $\rho$ is not just an arbitrary variable selected
because it is simple and leads to a consistent definition of community.  It has many theoretical properties which can be
compared to real communities, and helps in the formulation of a
consistent community detection framework.
In this section, we will derive various properties of $\rho$
which will prove useful in the exploration of the edge
density community definition.

\begin{figure}
  \centering
  \includegraphics[width=\halfwidth]{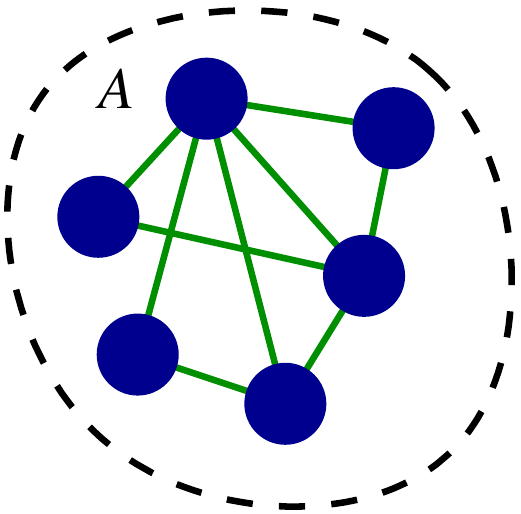}
  \caption[Illustration of average community edge
  density]{Illustration of average mean community edge
    density.  Community edge density $\rho_A \frac{9}{16}$ is equal to
    the mean of all node edge densities $\frac{1}{6}\left( \frac{5}{5}
      + \frac{2}{5} +
    \frac{4}{5} + \frac{3}{5} + \frac{2}{5} + \frac{2}{5} \right) =
  \left< \rho_{Aa_i} \right>$ This, and other similar invariants, are
  among the properties of edge density.}.
  \label{fig:indep-mean}
\end{figure}

Consider a community $A$ with $n_A$ members $a_i$
(Fig. \ref{fig:indep-mean}).  The edge density within $A$ is
$\rho_A$.  Each node $a_i$ has an edge density to the rest of the
community $\rho_{a_i}$.  The first property which we will show is that
the edge density of the community is equal to the average edge density
of the component nodes to the community,
\begin{equation}
  \label{eq:cmtyEdgeDensityAverageOfNodes}
  \rho_A = \left<\rho_{Aa_i}\right>,
\end{equation}
with the average taken over different community members $a_i$.  To
show this, we apply the edge density Eq.~\ref{eq:edgedensity} and
evaluate the average in Eq.~(\ref{eq:cmtyEdgeDensityAverageOfNodes}).  
This leads to 
\begin{eqnarray}
  \left<\rho_{Aa_i}\right> &=& {1\over n_A} \sum_a {e_{Aa} \over l_{Aa} }, \nonumber \\
                        &=& {e_A \over \frac{1}{2}n_A (n_A-1)}.
                        \label{eq:rho-community-mean}
\end{eqnarray}
In the above, the number of links for each node
is constant at $l_a = n_A-1$, and
${1 \over 2}n_A(n_A-1)$ is the number of links in a community of
$n_A$ nodes.

Next, when a node $x$ joins a community $A$ to form community $A\cup x$,
there is a relationship between the edge density of the combined unit
$\rho_{A\cup x}$ and that of the component edge densities $\rho_{A}$
and $\rho_{Ax}$,
\begin{eqnarray}
  \rho_{A\cup x} &=& \frac{e_{A\cup x}}{l_{A\cup x}} = \frac{e_A + e_{Ax}}{l_A + l_{Ax}}, \nonumber \\
    &=& \frac{l_A}{l_A + l_{Ax}} \rho_A + \frac{l_{Ax}} {l_A + l_{Ax}}\rho_{Ax}, \label{eq:rho-node-added-to-community}
\end{eqnarray}
which is simply the mean of $\rho_{A}$ and $\rho_{Ax}$ weighted by the
respective numbers of links $l_{A}$ and $l_{Ax}$.

A similar relationship can be shown for two non-overlapping
communities $A$ and $B$ with initial edge densities $\rho_A$ and
$\rho_B$ and a inter-community edge density $\rho_{AB}$.  When
two communities merge, the new edge density is the average of the edge
density of $A$, the edge density of $B$, and the inter-community edge
density $\rho_{AB}$, weighted by the corresponding number of links
$l_A$, $l_B$, and $l_{AB}$
\begin{eqnarray}
  \rho_{A\cup B} &=& \frac{l_A \rho_A + l_B \rho_B + l_{AB} \rho_{AB}}
                          {l_A+l_B+l_{AB}}.
\end{eqnarray}

These properties are useful for considering dynamics of community
detection algorithms.  For example, for a community $A$ and a node
$x$, if we know that $x$ should join $A$ ($\rho_{Ax}>\rho^*$), then
the final edge density of $A$ ($\rho_{A\cup x}$)can not decrease below
$\rho^*$.  Furthermore, if we have two communities $A$ and $B$ (with
$\rho_A>\rho^*$ and $\rho_B>\rho^*$), if the inter-community edge
density $\rho_{AB}>\rho^*$ (the criteria for community merging), then,
after merging, we are guaranteed that the new community $A \cup B$
must have $\rho_{A\cup B}$ satisfying the edge density condition
$\rho_{A\cup B} > \rho^*$.  These properties serve as a basic check
on the sensibility of our community definition.

When $\rho^* = 1$, then all communities must be fully connected (only
graph cliques are allowed as communities).  When $\rho^* > 1$,
communities can not exist as edge density can not be greater than one.
In this case, most CD methods will return ``communities'' which
actually consist of single
nodes, since there is never a case that multiple nodes can join
together.
When $\rho^* = 0$, there is no lower bound for community size, and all
nodes can collapse into one large community spanning the system,
however, this may not happen if the graph consists of disjoint subsets
of nodes and the exact dynamics of the CD process does not attempt to
join disjoint sets of nodes together.

\section{Model-dependent community properties}
\label{sec:ED-modeldependent}

The discussion of the
properties in the previous section makes one critical assumption:
when nodes are added to communities, they are previously
``unassigned,'' or in single-node zero-energy communities.  In these
cases, there is no energy barrier for removing nodes from the previous
community to which they are bound.  This is the case when constructing
local communities,
or when overlapping communities are allowed.
When this is not the case, in order to add a node $b$ to community
$A$, it must first be removed from some other community, say $B$.
Since the node $b$ is in community $B$, $b$ must have a binding energy to $B$ that must
first be overcome.  Any
energy released by moving $b$ to $A$ must first offset the energy
needed to remove $b$ from $B$.
In order to determine trade-offs between larger communities and greater edge
density, we must use the Hamiltonian from one of our set of models.
Towards this end, without loss of generality, we may use the APM.

\begin{figure}
  \centering
  \includegraphics[width=\halfwidth]{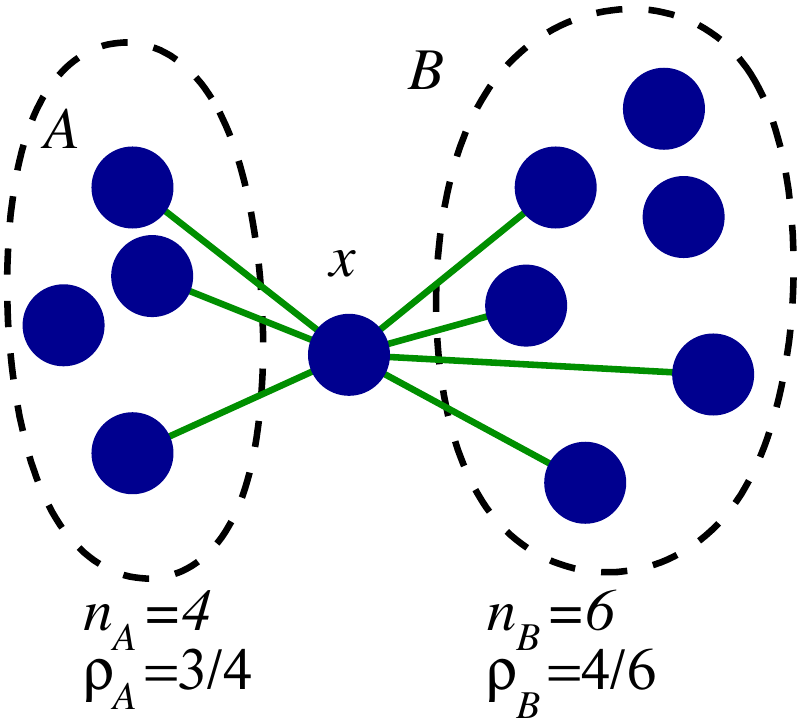}
  \caption[Edge density balance and node community membership]{Balance
    between placing a node $x$ into community $A$ or $B$.  In this
    plot, $x$ has a sufficient edge density to join either community,
    but when overlapping community assignments are not allowed, we can
    have $x$ join both.  The text discusses the criteria that edge
    density models use to assign $x$ to either $A$ or $B$.}
  \label{fig:edgedensity-ABx}
\end{figure}

In order to demonstrate the choices which the edge density models
make, we will use a simple thought experiment of one node $x$ which
can either join community $A$ or
$B$, as in Fig. \ref{fig:edgedensity-ABx}.  As a precondition , we must have
$\rho_{Ax}>\rho^*$ and
$\rho_{Bx}>\rho^*$, otherwise $x$ can not join both communities.
According to the edge density
community definition, with $\rho_{Ax}>\rho^*$ and $\rho_{Bx}>\rho^*$,
in isolation, or if overlapping community assignments were allowed, it
would be allowable for $x$ to join either
community, and this would be the preferred energy-minimizing move.
When overlaps are not allowed, $x$ must choose one of
$A$ or $B$ to join.

A similar situation occurs when node $b$, part of community $B$, has a
sufficient edge density to also join $A$.  The energy of addition to
$A$ must at least compensate
for the energy of removal of $b$ from $B$.  We imagine this in two parts:
first, the removal from $B$, and second, the choice between addition
to $A$ or $B$, allowing us to consider only the $A, B, x$ situation of
the previous paragraph without loss of generalization.

When using the VTPM, the node $x$ will choose
to join the community which most minimizes the energy.  It will join
$A$ when $E_{Ax} < E_{Bx}$, which can be rearranged to
\begin{eqnarray}
  n_A \left[ \rho_{Ax} - \rho^* \right]
    &>& n_B \left[ \rho_{Bx} - \rho^* \right]. \label{eq:apmBalance}
\end{eqnarray}
We see that $x$ will join the community $A$ or $B$ to which it has the
largest
excess edge density $\rho_{Cx} - \rho^*$, weighted by the community
size $n_A$ or $n_B$.  We note the following: (a) we assume $\rho^* < 1$.  The
cases where $\rho \ge 1$ are discussed above. (b) we assume $\rho_{Ax} >
\rho^*$ and $\rho_{Bx} > \rho^*$.  If both are less than $\rho^*$, $x$
can join neither community, if one is less than $\rho^*$, we see the node
will join the other community (c) If $\rho^* = 0$, then
$x$ will join the community of greater size.  (d) If $\rho_{Ax} =
\rho_{Bx}$, $x$ will join the community of greater size.  (e) if $n_A
= n_B$, according to Eq.~(\ref{eq:apmBalance}), $x$ will join the
community to which it has greater $\rho$.

We see that these results support the idea that a node, when faced
with other communities of equal sizes, will join the community to
which it shares the greatest edge density, however, there is also a
competing preference towards smaller, more dense, communities.

It deserves emphasis that the results from this section are derived
for one particular instance of the edge density community definition,
the one derived from the VTPM.  For unweighted graphs, this model will give equivalent results to the APM.  The APM has been shown to
be an effective community detection method in a wide
variety of situations \cite{ronhovde2010local,
  ronhovde2009multiresolution, lancichinetti2009community}.

\section{Simple edge density community detection algorithms}
\label{sec:ED-algorithms}

The purpose of this paper is not to discuss specific edge density
community detection algorithms, instead, we focus on the
\emph{ground-state} properties of this community definition.  There
are various published algorithms which can be used directly with edge
densities.  Previous work from our group used a global algorithm,
beginning with every node in a different community and merging
communities until some energy minimum is
reached\cite{ronhovde2009multiresolution, ronhovde2010local,
  hu2011phase, hu2011replica}.  Alternatively, there are various local
methods, which build up a single community around single
nodes\cite{lancichinetti2009detecting, lancichinetti2011finding,
  lee2010detecting}.

Unlike other methods, there is no heuristic for determining the proper
$\rho^*$.  Instead, we use a multi-replica inference method where
independent ``replicas'' are solved at the same value of $\rho^*$, and
their similarity is compared.  If, for a given value of $\rho^*$, the
independent replicas minimize to give similar community structures,
this is considered to be a ``good'' value of $\rho^*$.  This has
proven to be a reliable and robust method for community detection.
For addition information, see Sec.~\ref{sec:ED-multireplica}.

Further information and considerations about these methods is found in
Appendix~\ref{sec:ED-app-algorithms}.

\section{Edge densities in real networks}
\label{sec:ED-edgedensityscaling}

Our edge density definition assumes that $\rho^*$ is a useful
resolution parameter, which can select for larger or smaller
communities.  We have operated on an intuition that a larger $\rho^*$
selects for a smaller, more densely connected communities, while a
smaller $\rho^*$ selects larger, less densely connected communities.
In this section, we will directly consider this assumption.
Recently, Yang and Leskovec (YL) have taken a variety of
real networks, mainly social networks, and rigorously studied their
properties with respect to size, overlap regions, and other
parameters\cite{yang2012structure}.  They find that the number of
edges within a community tends to grow with a power in the range $(1,
2)$, with observed values of $1.1$ and $1.5$.  For some exponent value
$\nu$, it is thus found
\begin{equation}
  e \propto n^{\nu}.
\end{equation}
However, maximal number of edges ($l$ in our nomenclature) grows as
\begin{equation}
  l = \frac{1}{2} n (n-1) \propto n^2,
\end{equation}
thus we find a scaling relation of edge density of
\begin{equation}
  \rho = \frac{e}{l} \propto \frac{n^{\nu}}{n^2} \propto n^{\nu-2}.
\end{equation}
As long as $\nu\in (1,2)$, we find $\rho$ decreases as $n$ increases.
For example, YL observed social networks to have an exponent value
of $\nu\approx1.5$, giving us
\begin{equation}
  \rho \propto n^{-.5}.
\end{equation}
As we see, this validates one of the central tenets of the edge
density community definition: as communities grow larger, the edge
density tends to decrease.  By specifying a $\rho^*$, we implicitly
specify a size scale of community which we will then detect.  Another
way to view this is from the standpoint of an agglomorative community
detection alogrithm.  Starting from a dense core, as each additional
node is added to a given community, the community edge density on
average decreases with each additional node, lowering the edge
density.  As we keep adding nodes to the community, eventually, the
community will grow so large that adding extra nodes will decrease
$\rho$ below $\rho^*$.

\section{Overlapping communities}
\label{sec:ED-overlaps}

Many networks have community structure which can most naturally be
described as \emph{overlapping}.  In this viewpoint, there are certain
nodes which can be reasonably included in multiple groups.  Perhaps
the most standard example of this situation is social networks: any
one person will be involved in groups corresponding to work, family,
hobbies, etc.  The overlaps can consist of single nodes or larger
subsets.  Not all community detection methods can be extended to
handle overlapping nodes.  For example, the Newman betweenness
algorithm progressively cuts edges until modularity maximization
states that final communities are found \cite{newman2004finding}.
Because each cut is final, there is no ability to create overlapping
communities.  Different methods, such as clique percolation
\cite{palla2005uncovering}, 
attempt to detect overlapping community structure.  There are a
variety of local community definitions, including the
community-centric Potts model formulations of Eq.~(\ref{eq:APMcmty}),
can detect overlapping communities by virtue of \emph{independently}
detecting each community\cite{lancichinetti2009detecting}.

The work of YL also looked at the characteristic of overlapping
regions of communities\cite{yang2012structure}.
According to the behavior seen by YL, regions of overlap between
communities have edge density contributions from both
communities\cite{yang2012structure,
  simmel1964conflict,feld1981focused}.  Thus, these overlap regions have a
greater density than the individual communities.  It had previously
been assumed that these regions of overlapping communities had a
smaller edge density than either of the non-overlapping regions, and past
methods of community detection make the opposite assumption, and thus
not suitable for community detection with the observed behavior
\cite{palla2005uncovering,ahn2010link, lancichinetti2009benchmarks}.
We can show directly that the edge density community definition
properly handles this case.

\begin{figure}
  \centering
  \includegraphics[width=5cm]{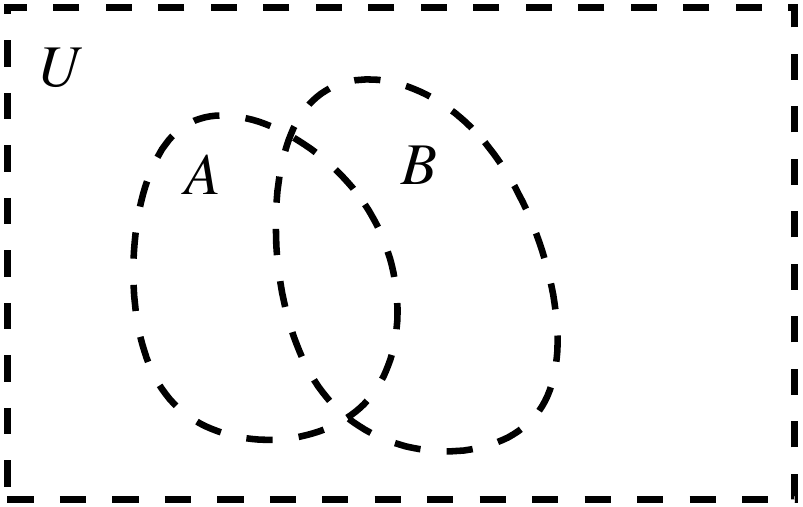}
  \begin{tabular}[b]{c|c|c|c|c}
                & $A-B$ & $B-A$ & $A\cap B$ & $U-A-B$  \\ \hline
     $A-B$      & .5    &  0    &  .5             & 0  \\ \hline
     $B-A$      & 0     &  .5   &  .5             & 0  \\ \hline
     $A\cap B$  & .5    &  .5   &  .75            & 0  \\ \hline
     $U-A-B$    & 0     &  0    &  0              & 0
  \end{tabular}
  \caption[Edge density and overlapping regions]{Schematic of two
    communities $A$ and $B$ with $p_A=p_B=.5$, and their overlap edge
    densities.  This illustrates the affiliation graph model edge
    densities between each of the distinct regions of the figures,
    with a ``dense overlap'' in the $A\cap B$ region, as determined by
    YL \cite{yang2012structure}.  According to YL, currently existing
    community definition models can not detect these communities.  The
    table indicates the probability for an edge being present between
    any two regions of the plot: the ``$-$'' operator indicates set
    difference, $A-B$ indicates the region of $A$ excluding $B$.}
  \label{fig:edgedensityoverlap}
\end{figure}

Let us look at a diagram of two overlapping communities $A$ and $B$,
embedded in a universe of nodes (Fig \ref{fig:edgedensityoverlap}, upper).  We denote the universe of nodes $U$,
the two communities $A$ and $B$, the region of community $A$ excluding
$B$ as $A-B$ and vice versa, the overlap region $A\cap B$, and the
universe of nodes excluding either community, $U-A-B$.
Fig.~\ref{fig:edgedensityoverlap} is an schematic of this
situation.  Communities $A$ and $B$ both have internal edge densities
of .5.  The overlapping region $A \cap B$ has an edge density of
.75.  The edge density between $A - B$ and $A \cap B$ is .5, and vice
versa for $B$ and $A$.

With $\rho^*=.75$, only the overlapping region $A \cap B$ will be
detected, with $\rho^*=.5$, communities will expand to cover the
region $A$ or the region $B$, but not $A \cup B$.  The reason for
this is that, if we have currently detected the community $A$, any one
one node $x$ in $B-A$ will have a $p=.5$ edge density to $A \cap
B$, but only $p=.5$ edge density to $A - B$, thus the average
connection probability to the entire set $A$ is $p_{Ax} \frac{.05 n_{A-B} +
  .5 n_{A\cap B} }{n_A} < .5$, thus, $p_{Ax} < \rho^*=.5$.  This
is the desired behavior.  The same will be true of a node $y$ in $A-B$
attempting to join community $B$.  Thus, by applying the edge density
community definition, we can get exactly the desired behavior:
for $\rho^*=.5$, we detect $A$ and $B$, while for $\rho^*=.75$, we detect
$A\cap B$.

\section{The Affiliation Graph Model}
\label{sec:ED-AGM}

In order to model the properties they observed, YL described the
``Affiliation Graph Model'' (AGM)\cite{yang2012structure}.  This model
is similar to stochastic block models in that edges are drawn in only
based on the community memberships of the two nodes.  However, nodes
are allowed to belong to multiple communities, and the communities are
not restricted to being disjoint.  This can be interpreted as a
bipartite graph from ``people'' to shared ``affiliations''.  In the
AGM itself, the affiliations are not present, and instead each shared
affiliation incorporates chance of shared edge between ``people''
(nodes in the graph).  A similar concept of affiliations grouping has
been considered before in a sociological and network
context\cite{feld1981focused,breiger1974duality}, and such work has
hinted that under certain generation processes, such models could
produce power-law distribution of node
degrees\cite{yang2012structure}.  Because this benchmark model
incorporates dense overlap, YL claim that current community detection
methods are not able to successfully detect communities in this type
of graph.

The basic idea behind AGMs is that communities (``affiliations'') are
chosen from a universe of nodes.  These could be non-overlapping and
spanning the universe (yielding a \emph{planted l-partition}, or
Stochastic Block Model), or any assortment of hierarchical,
overlapping, subset-containing, or other arrangements.  Then, we add
an edge with probability $p$ for \emph{each} shared affiliation.
Since there can be multiple shared communities per node, we allow each
shared affiliation a chance to produce an edge.  This gives us an
ultimate edge probability between nodes $x$ and $y$ of
\begin{equation}
  p_{xy} = 1 - \prod_{C} (1-p_C) \label{eq:AGMprobabilities}
\end{equation}
where the product is over all communities $C$ which contains both nodes
$x$ and $y$.  This is the complement of the probability that none of the
shared affiliations produce a community.

\begin{figure}
  \centering
  \begin{tabular}{cc}
    (a) SBM & (b) AGM \\
    \includegraphics[width=\halfwidth]{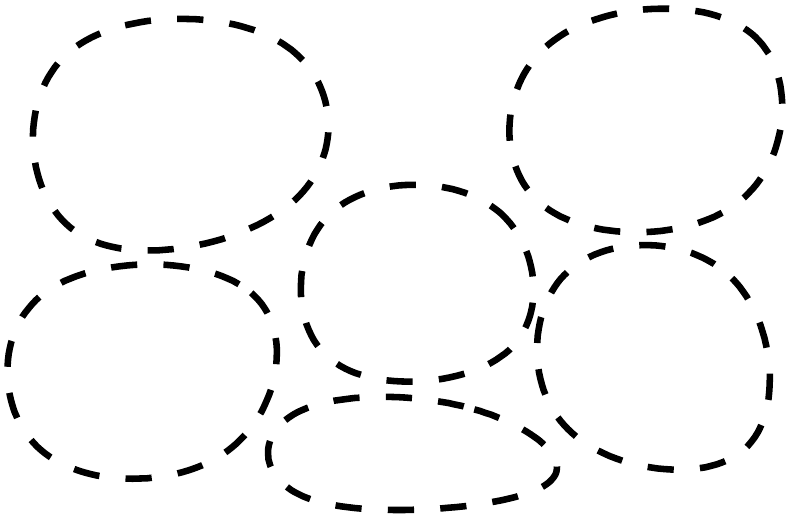} &
    \includegraphics[width=\halfwidth]{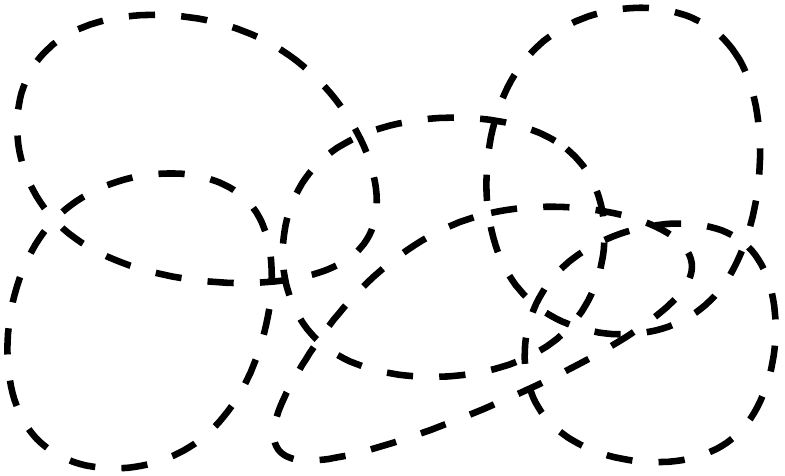}
  \end{tabular}
  \caption[Stochastic block models vs affiliation graph
  models]{Schematic of stochastic block modes (SBM) vs affiliation
    grapm model (AGM) graphs.  (a) The SBM graphs are non-overlapping,
    while (b) the AGM graphs allow overlaps.  Note that in our
    specific instances of AGM graphs, every node is it at least one
    community.}
  \label{fig:SBMvsAGM}
\end{figure}

The instance of AGM networks we study here is
constructed as such: A universe of $N$ nodes is taken.  We
take $q$ communities of $n$ nodes each, with $nq \geq N$, in two
steps: first, we initialize the communities with non-overlapping
communities of $N/q \le n$ nodes.  Then, for each community, we add
additional nodes necessary to make $n$ nodes per community by randomly
choosing from all other nodes.  There are no restrictions to the
maximum number of communities to which a node can belong.  The edge
probability between any two nodes is given by Eq.~(\ref{eq:AGMprobabilities}).
Our particular AGM graphs can be
identified by the parameters $(q, n, p)$.
The AGM benchmark graphs differ from stochastic block model graphs by
allowing overlaps, Fig. \ref{fig:SBMvsAGM}.

\begin{figure}
  \centering
  \begin{tabular}{cc}
    (a) Hierarchical & (b) Multi-layer \\
    \includegraphics[width=\halfwidth]{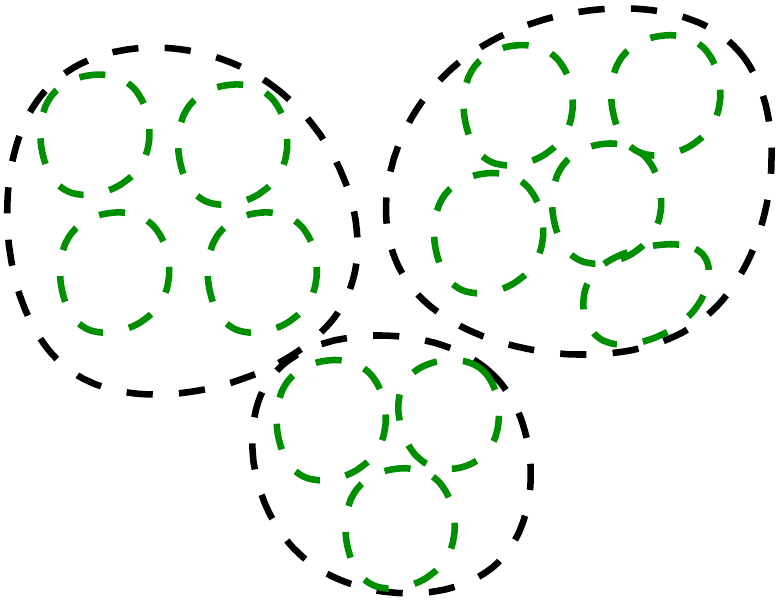} &
    \includegraphics[width=\halfwidth]{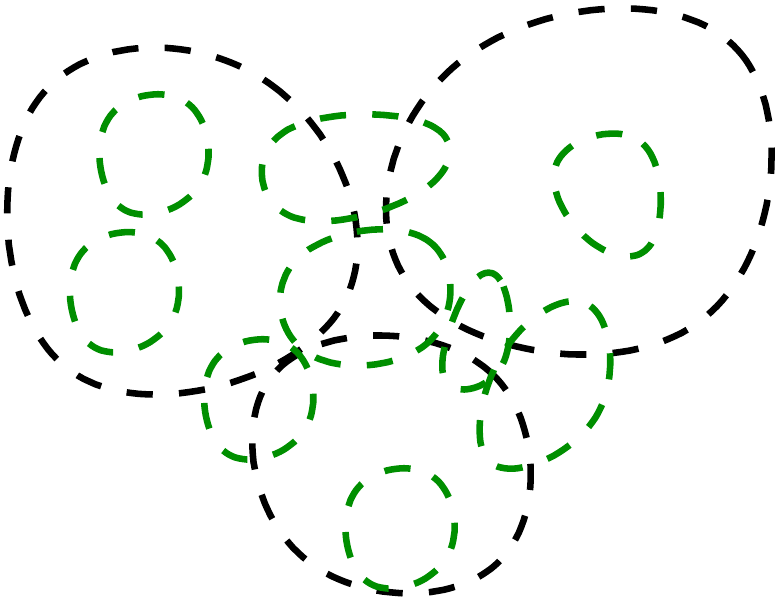}
  \end{tabular}
  \caption[Hierarchical vs multi-layer community structure]{Schematic
    of hierarchical vs multi-layer.  (a) The hierarchical
    graphs have small communities (smaller circles, green) contained entirely
    within large communities.  (b) The multi-layer graphs have small
    communities which are allowed to span multiple large communities.
    In our specific instances of AGM graphs, every node is it at least one
    community, and small communities are assigned
    \emph{completely} independently of large communities, allowing
    extremely high degrees of mixing between small and large
    communities.  There have been no proposals of this type of
    multi-layer community detection in the literature, but we
    will demonstrate our methods detecting these communities
    properly.}
  \label{fig:HierarchicalVsMultilayer}
\end{figure}

Further, we may produce \emph{multi-layer} benchmark graphs.  It is
traditional to produce hierarchical graphs for consideration, where
small communities are contained within single large communities.  To
create our multi-layer graphs, we create AGM graphs
\emph{independently} over the same universe of nodes, and then
merge the set of edges.  There is no requirement
that each small community be contained within a single large
community, and in fact, for each community, each small
community is likely to overlap with \emph{every} large community.  The
amount of overlap present here is unparalleled in the existing literature.
Fig. \ref{fig:HierarchicalVsMultilayer} illustrates the difference
between hierarchical and multi-layer community assignments.  To
produce multi-layer AGMs, we take two AGMs independently from the
universe of nodes and overlay edges.  Edge probabilities are still
calculated via Eq.~(\ref{eq:AGMprobabilities}), but now the sum over
communities $C$ contains multiple layers of communities.  We denote
our multi-layer AGMs via the nomenclature $((q_1, n_1, p_1), (q_2,
n_2, p_2))$.  Our final graphs are thus a combination of the overlaps
of Fig. \ref{fig:SBMvsAGM} and multi-layer character of
Fig. \ref{fig:HierarchicalVsMultilayer}.

We now examine actual results from application of the absolute Potts
model and variable topology Potts model to our overlapping
community benchmarks using the multi-replica inference framework of
Sec.~\ref{sec:ED-multireplica}.  We use the F-score measures of
Appendix \ref{sec:thesis-F1} to judge the performance of our
algorithms.  We study both inter-replica
$F_1^{IR}$, which is used to identify the correct $\rho^*$, and
$F_1^0$, a measure of how well
we detected the planted communities.  According to the canonical
multiresolution algorithm, extrema of uniformity between replicas
($F_1^{IR}$)
indicate $\rho^*$ (or $\gamma$) values likely to be significant.  $F_1^0$ indicates how well we detect the planted communities.  In practice, we want a maximum of $F_1^{IR}$ to correspond to the maximum of $F_1^0$.
When both of these measures to peak at the same value of $\rho^*$
to indicate that we can recover the
planted states and that we would be able to infer the correct value of
$\rho^*$ if the planted states were not known \emph{a priori}.

\begin{figure}
  \centering
  \begin{tabular}{cc}
    (a) & (b) \\
    \includegraphics[width=\halfwidth]{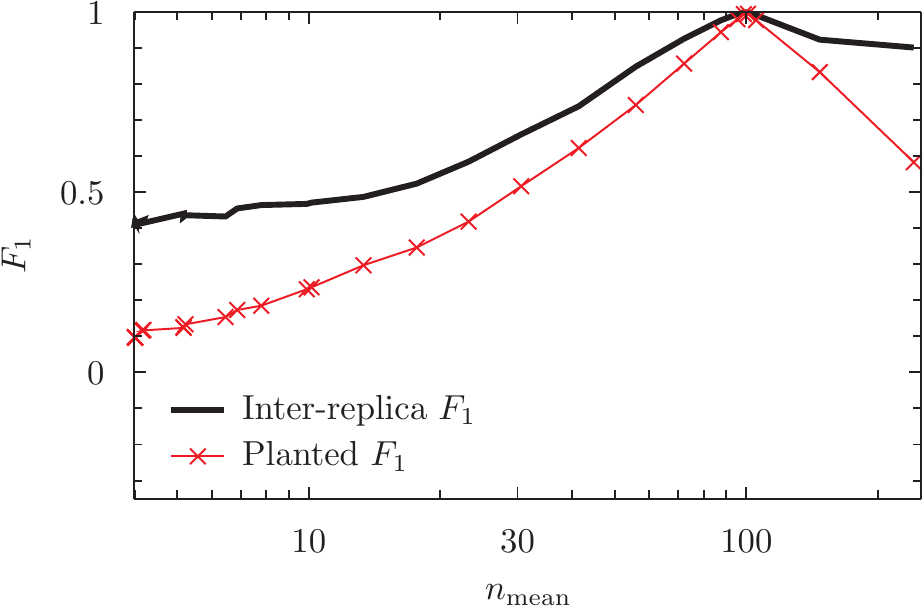} &
    \includegraphics[width=\halfwidth]{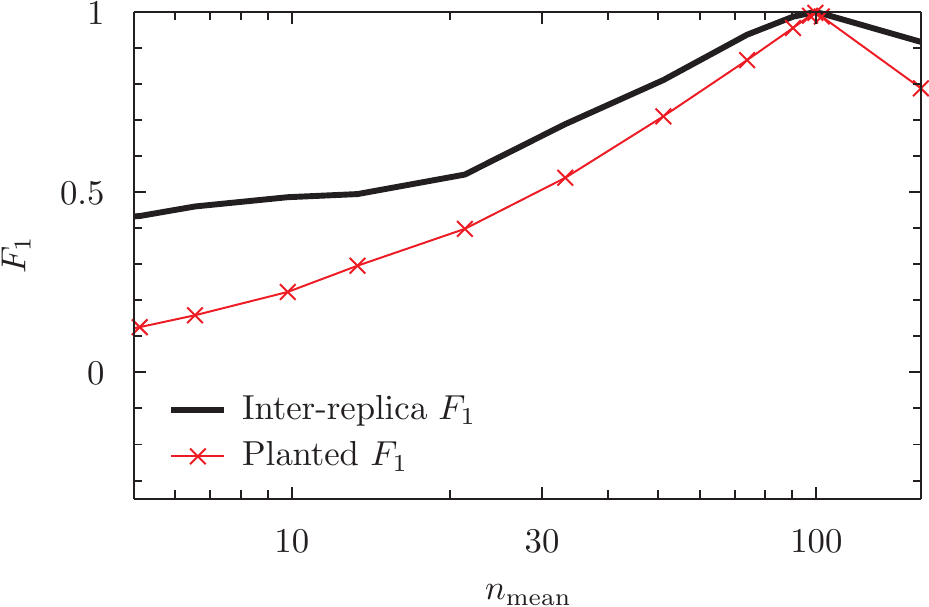} \\
    \includegraphics[width=\halfwidth]{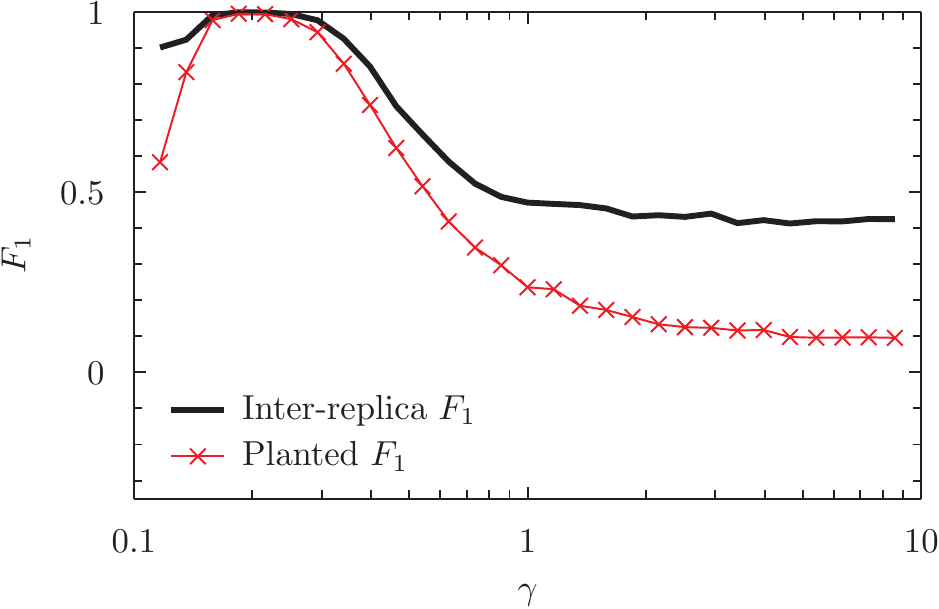} &
    \includegraphics[width=\halfwidth]{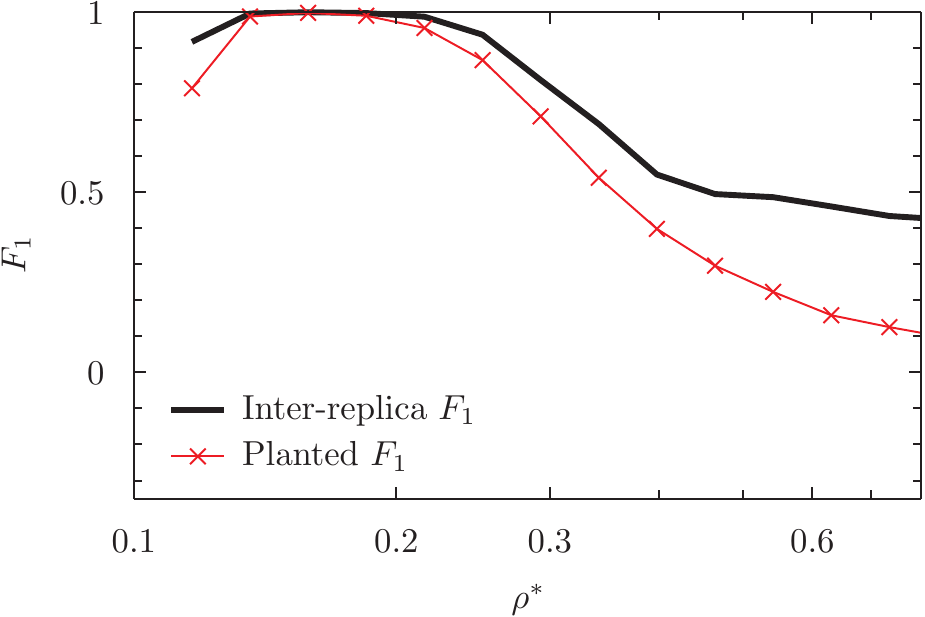}
  \end{tabular}
  \caption[Detection of communities on single-layer affiliation graph
  model graphs]{Community detection results on single-layer AGM graph with
    parameters number of communities $q=20$, number of nodes per community $n=100$, and edge probability $p=.25$ in a total universe of $N=1000$ nodes, showing perfect community
    detection.
    $F_1^0$ indicates our ability to detect the known community
    structure; a value of unity indicates perfect community
    detection.  $F_1^{IR}$ indicates our ability to locate the correct
    resolutions (value of $\rho^*$) without \textit{a priori} information
    about community sizes.  When both measures becoming unity at the same
    $\rho^*$ value, our methods can both infer the correct
    resolutions and achieves perfect detection at that resolution.
    Similar results are achievable with a wide range of
    AGM parameters, demonstrating robustness of the methods.  Upper plots show community detection with respect
    to $n_\mathrm{mean}$, showing that the methods do indeed detect the correct community size $n=100$.  The column (a) uses the absolute Potts model, and the
    column (b) uses the variable topology Potts models.  The lower plots
    show the role of the respective scale parameter $\gamma$ for the
    APM and $\rho^*$ for the VTPM. }
  \label{fig:overlap1layer}
\end{figure}

\begin{figure}
  \centering
  \begin{tabular}{cc}
    (a) & (b) \\
    \includegraphics[width=\halfwidth]{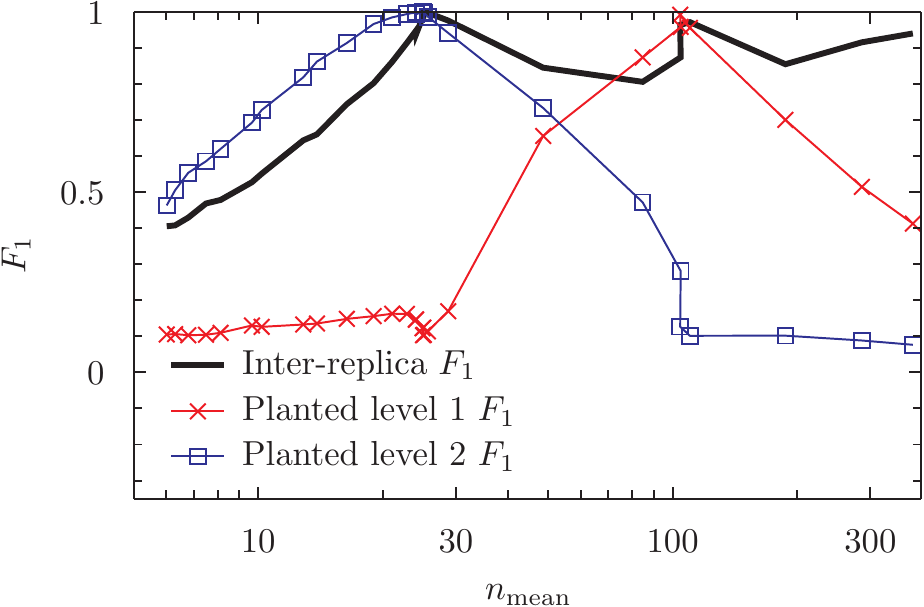} &
    \includegraphics[width=\halfwidth]{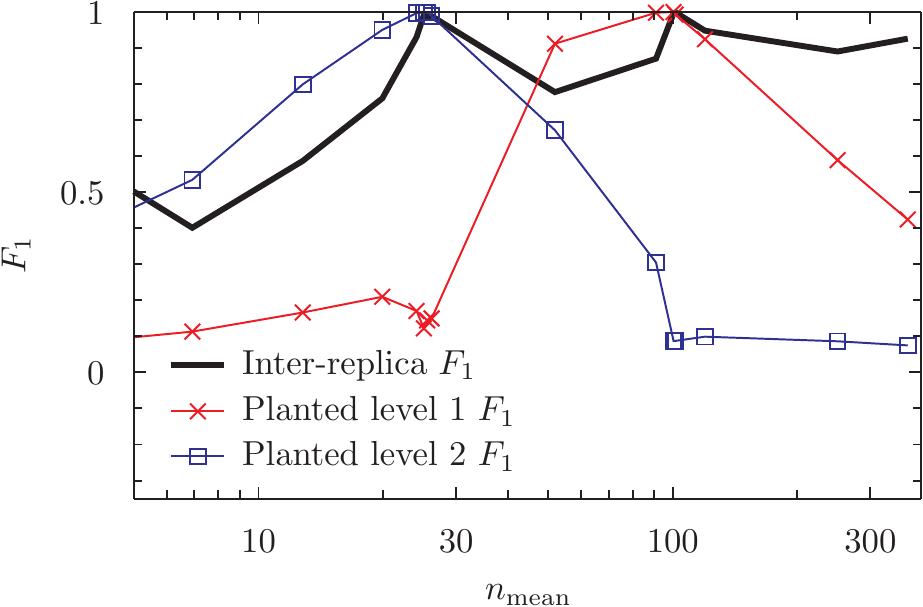} \\
    \includegraphics[width=\halfwidth]{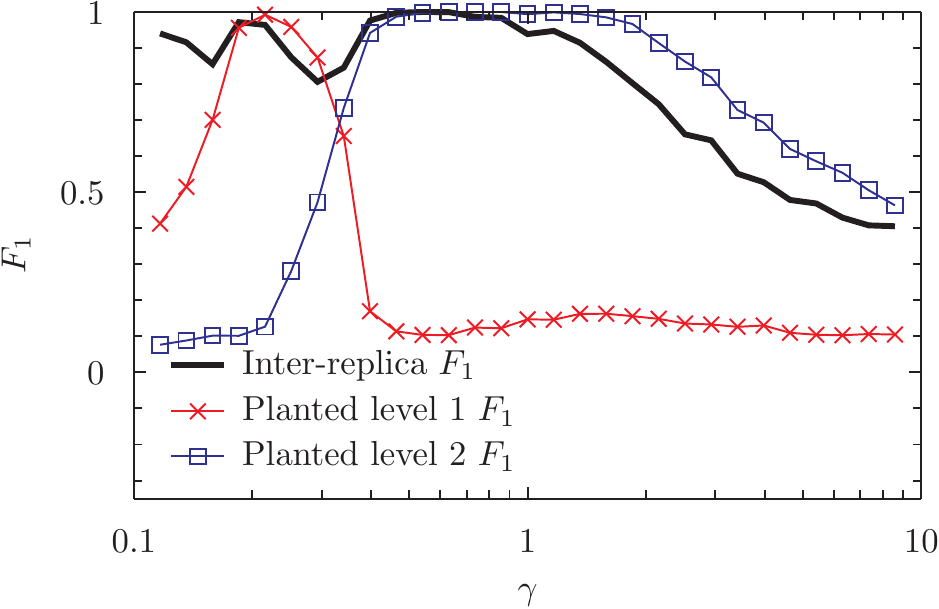} &
    \includegraphics[width=\halfwidth]{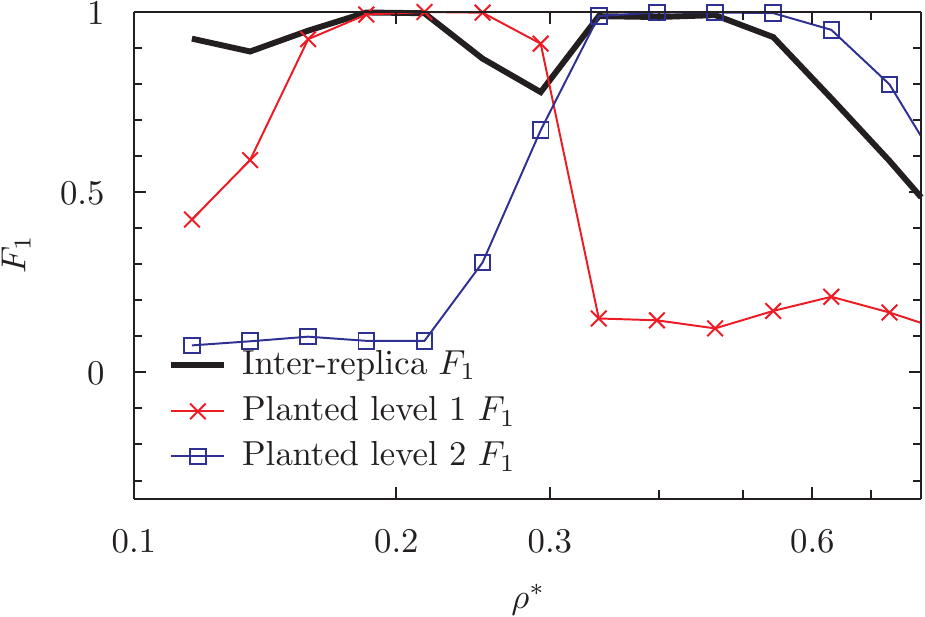}
  \end{tabular}
  \caption[Detection of communities on multi-layer affiliation graph
  model graphs]{
    Community detection results on multi-layer AGM graphs.

Multi-layer AGM graphs with parameters (a) $((q_1=16,
    n_1=100, p_1=.3), (q_2=64, n_2=25, p_2=.75))$ and (b) $((q_1=16,
    n_1=100, p_1=.35), (q_2=64, n_2=25, p_2=.75))$.  Both
    inter-replica and known detection measures are plotted as in
    Fig.~\ref{fig:overlap1layer}.
    Note that these
    parameters are ``cherry-picked'' to give good results; successful
    community detection is not possible for arbitrary parameters.  At
    arbitrary parameters, one layer is perfectly detectable and one
    layer is noticeable but without perfect detection.
    Upper/lower rows and left/right columns maintain the same interpretation as in Fig.\ref{fig:overlap1layer}
  }
  \label{fig:overlap2layer}
\end{figure}

Fig. \ref{fig:overlap1layer} shows results for single-layer AGM
graphs.  We see that we can perfectly recover communities for both
situations, and are able to infer the correct value of $\rho^*$ (or
APM $\gamma$) without \emph{a priori} knowledge.  We see that we are
not only able to detect the planted communities (planted $F_1=1$), but
to know where that would be (inter-replica $F_1=1$).
Fig. \ref{fig:overlap2layer} shows results from application to
multi-layer AGM graphs.  We see accuracy comparable to single-layer
graphs, with slightly less well defined peaks.  It should be noted, we
can not solve this for arbitrary parameters.  The $(q,n,p)$ of the two
layers are hand-chosen to give good results by adjust the
probabilities of large and small communities.  Because of this, the
methods here would not necessarily be useful for real graphs.  We
choose this example to demonstrate the power and limitations of the
method.  It is unlikely that real graphs would be as complex to solve
as the multi-layer fully independent random case.

In a future work, we will systematically apply edge density methods to
a wider variety of AGMs, demonstrating the usefulness of edge density
methods.  The models we will consider will include various
combinations of power law community sizes, different forms of
hierarchy, communities within communities, homeless nodes, and more.
In addition, we will provide theory information regarding the limits
of edge density methods in the face of large amounts of dense overlap.

\section{Weighted graphs}
\label{sec:ED-weighted}

One area of modern community detection research is the
subject of weighted graphs.  In weighted graphs, not every edge has
equal importance.  Any modern community detection method should be
able to handle weighted graphs.
A high weight
will indicate an edge which plays a major role in the graph, while a
low weight indicates an edge which does not significantly affect the
graph.  An unweighted graph can be considered a weighted graph of
edges of weight one, so by convention, a weight of zero corresponds to
an edge which is not present, and a weight of one corresponds to an
unweighted edge.

In order to adapt edge density to weighted graphs, we make a
simple substitution.  We consider the number of edges to be
the sum of weights of all edges $i$,
\begin{equation}
  e = \sum_i w_i
\end{equation}
with the sum taken over all edges.  Thus, the edge density for any
grouping of nodes $X$ is
\begin{equation}
\label{rox}
  \rho_{X} = \sum_{i \in X} \frac{w_i}{l_X}.
\end{equation}
The grouping $X$ can be that of any community $A$, a pair of communities $AB$, or
any of the other situations depicted in Fig.~\ref{fig:vardefs}.  As to be expected
for any reasonable extension of its counterpart of the unweighted graphs, 
we note that, indeed, for unweighted graphs Eq. (\ref{rox})
reduces to the same definition appearing in Eq.~(\ref{eq:edgedensity}).
Furthermore, edges can be weighted greater than one for very important
edges.  Adversarial edges (edges which favor being in different
communities) can be weighted less than zero.  All of our analysis in the previous
sections remains valid \emph{mutatis mutandis}, with the caveat that $\rho$ is no longer
limited to the range $[0,1]$.  The edge density can exceed unity when
there are many edges with weight greater than one, or less than zero
if there are enough adversarial edges.  Furthermore, we maintain a
property of linearity of edge densities.  If linearity is not desired,
a power (or other mapping function) could be applied to edge densities
before summing to get the ``number of edges'' stand-in.

The APM Hamiltonian, Eq.~(\ref{eq:APMcmty}), can handle weighted
graphs, with $A_{aa'}$ being a matrix of edge weights for present edges
and $B_{aa'}=1$ if there is no edge.  There are two caveats.  First,
the APM model
uses the number of missing edges, for which we use $l - e$ in
Eq.~(\ref{eq:APMcmty2}) (the
maximal number of possible edges, minus the number of actual edges).
However, when $e = \sum w_i$, this is no longer necessarily true: $e$
no longer is identical with the number of existing edges.
Thus, Eq.~(\ref{eq:APMHamiltonian}) no longer corresponds to Eqs.~(\ref{eq:APMspin}, \ref{eq:APMcmty})
Second, in our final density formulation of the APM Hamiltonian
(Eq.~\ref{eq:APMHamiltonian}), there is a factor of $\frac{1}{1-\rho^*}$.
This becomes zero or negative when $\rho^* \ge 1$, thus rendering our
derivations invalid.

The VTPM avoids both of these limitations, and allows a natural
extension to weighted graphs with no discontinuity for $\rho<0$ or
$\rho>1$.  We reiterate that as we have shown that the APM and VTPM are
  identical for unweighted graphs, we know, even without dedicated
  experimentation, that the VTPM is a successful community detection
  method for a broader class of graphs.

\section{Extensions}
\label{sec:ED-directedetc}

\emph{Multigraphs} are graphs that allow more than one edge to be
between any pair of nodes.  As we sum over all edges, multigraphs can
be very naturally included in the summation over edges.  However, note that
under this formulation, a multigraph is seen as equivalent to a
weighted graph, with each edge having a weight equal to the sum of the
weights of the other edges.  If this procedure loses essential
information about the graph, a different method will be needed.
Perhaps multiple edges could be combined into one with a different
weighting function, however, without a rigorous analysis of the most
important aspects of multigraphs, we can only speculate.  For now, all
we do is point out that edge density is not incompatible with the
concept of multigraphs.

If our graphs have multiple types of edges, or multiple distinct forms
of weight per edge $w_0, w_1, \ldots$ generalized to a vector
$\mathbf{\hat{w}}$, we can generalize the total edge weight as a
weighted average $w = \mathbf{\hat{w}}\cdot \mathbf{\hat{u}}$, with
$\mathbf{\hat{u}}$ being a unit vector weighting the different
individual weight contributions.  This allows us to use a different
combination of component weights depending on our intended community
detection goals.

The methods presented in this paper use only edges, and not higher
order correlations such as triangles (3-cliques).  In order to use
e.g. triangles, we would define a ``triangle density'' as
\begin{equation}
  \rho_C^t = \frac{t_C}{\frac{1}{6}n_C(n_C-1)(n_C-2)}
\end{equation}
with $t_C$ being the number of triangles in community $C$, $n_C$ being
the number of nodes in community $CC$, and the denominator being the
maximal possible number of triangles in community $C$.  This could be
extended to a number of other higher-order correlations of any
structural motif rather trivially.  Many real networks have been shown
to contain non-trivial patterns in these higher order correlations,
which are for the most part not presently considered in community
detection methods\cite{newman2003structure}.

\section{Discussion}
\label{sec:ED-limitations}\label{sec:ED-discussion}

Despite edge
density well describing many benchmarks and real graphs in an
effective manner, our definitions and protocols have
certain limitations.  While these limitations exist, past use of
edge density methods has shown they are not barriers for most
important problems.  The edge
density community definition is \emph{simple}.  Thus, we expect the edge density community
definition to be general and easily extendable.

For large sparse networks, the total number of edges is proportional
to the number of nodes ($e \propto cn$, $c=O(1)$), while the number of
possible edges is $l \propto p\frac{1}{2}n(n-1)$.  The edge densities
are driven to zero exists as the number of nodes per community $n$
increases,
\begin{equation}
  \rho
  \propto \frac{e}{l}
  \propto \frac{c n}{p\frac{1}{2}n(n-1)}
  \propto \frac{c}{pn}
  \to 0.
\end{equation}
When all edge densities become small, it become progressively more
difficult for the parameter $\rho^*$ to distinguish communities.  This
general problem is discussed in more general terms in a companion work
\cite{darst2013noise}.  Overall and
inter-community edge density decreases with $N$ (total number of
nodes), while intra-community edge density decreases with $n$, the
number of nodes per community.  However, in the case of somewhat
fixed-sized communities, the inter-community edge density is driven
low much faster than intra-community edge density, allowing $\rho$ to
still be successfully used to distinguish the communities.  The
success of edge density community definition will depend on the
precise graph and ``ground-truth'' community characteristics, and can
be studied over various classes of graphs.

Studies of real networks have shown a power law distribution of
community
sizes\cite{palla2005uncovering,guimera2003self,clauset2004finding}.  A
variation of community size does not affect
the edge density community definition, as long as the community has a
uniform $p$ of each edge existing, the edge density community
definition will be able to properly detect its communities.

Studies have also shown that real networks have power-law
distributions of node
degrees\cite{newman2003structure,boccaletti2006complex,albert1999internet,
  barabasi1999emergence}. If, upon further analysis, this power law
degree distribution corresponds to a power law distribution of
\emph{internal} degrees (number of edges connecting to other nodes
inside a community), then the distribution of node ($a$) to community ($A$)
densities $\rho_{Aa}=e_{Aa}/(n_A-1)$ will then be power law
distributed as well.
The edge density community definition can still be relevant
to these graphs.  In these cases, $\rho^*$ will determine the edge
density for the lowest-degree nodes of the community.  The high-degree
nodes will still be included as part of the community to which they
have the greatest connecting edge density.  If a high-degree node has many
of its edges spread out among other communities, it will not be
misclassified as long as its node-to-correct-community edge density is
greater than
$\rho^*$ and it shares the greatest edge density with the correct
community.
The work of Lattanzi and Sivakuma \cite{lattanzi2009affiliation} provides another possibility which
will be investigated in a future work.  Using a model similar to AGMs,
they have shown a natural appearance of power law behavior in
affiliation graphs.  If this holds for AGMs, then power law node
degrees may be consistent with constant internal densities - rendering
our current work indeed relevant.

According to the simple definitions listed here, \emph{every} node
must have an edge density of greater than $\rho^*$ to its community.
This appears to be a fairly heavy restriction: if there is a node with
less than $\rho^*$ edge density to \emph{any} other community, it will
be forever alone.  One method of working around this would be to then
allow isolated nodes to join whatever community they have the greatest
edge density connection with.  Alternative schemes could be
developed, where only the average community edge density must be
greater than $\rho^*$, and certain individual nodes can have an edge
density of less than $\rho^*$.  Regardless, internal edge density is
only capable of detecting assortative communities, where nodes are
connected to similar nodes.

In reality, no single community definition is expected to work across
all classes of graphs.  Edge densities may describe graphs arising
from a certain generation process (such as shared affiliation social
networks), information flows may describe data-centric networks, and
other methods may describe scientific citation networks, among many
possibilities.  Certain community definitions should not be studied at
the exclusion of others, and it is important to understand all
community definitions to know the realm of their applicability.
Furthermore, rigorously understanding the resulting communities from a
variety of community detection algorithms provides a tantalizing
inverse-community detection possibility.  Suppose we had a graph which
an unknown complex structure, but we \emph{did} know ground-truth
communities.  By applying community detections algorithms to this
graph, and comparing the returned communities to the planted
communities, we may learn something about the graph generation process
itself.

\section{Avoiding over-optimization}
\label{sec:ED-overoptimization}

As stated in Sec.~\ref{sec:ED-binomial}, the edge density variable
$\rho$ is directly analogous to the probability $p$ used to generate
many common benchmarks.  Since our detection method exactly matches
the creation processes of common benchmark graphs, it is easy to
understand why we achieve such accurate detection.  Our definition
will be useful for the cases where real graphs are generated in an
analogous manner.  A current topic of research is the \emph{processes}
which generate various real-world networks; for example, processes
have been proposed giving rise to power law degree distributions, and
also with constant edge probability, as in AGMs.  Once more is known
about these processes, we can better understand what community
definitions are optimal.  In addition, there is no reason to believe
that one community definition will be optimal in all types of graphs, and
we must have many available techniques in our toolbox to be able to
respond to whatever problems may occur.

Nevertheless, many current benchmark graphs \emph{are} generated using
edge densities, and can be described as such.  Our companion work
expands on this, and uses edge density - the lowest common denominator
of many methods, to learn about the structural source of community
detection limits in stochastic block model graphs.  Furthermore, due
to the highly symmetric nature of these equal-sized stochastic block
model graphs, our edge density results generalize to almost all
community detection methods.

\section{Conclusions}
\label{sec:ED-conclusions}

Edge density community detection methods have been used before, but
despite this fact, the underlying theory of such methods have not
been explicitly defined and fully explored.  This work fills this
gap, and in the process formally expands the edge density community
definition to include
important new classes of graphs, such as weighted graphs,
and the possibility for overlapping communities.  One of the most
important features of our edge density community definition is that it
\emph{simple}.  One equation, $\rho=e/l$ (Eq.~(\ref{eq:edgedensity})), is all that is needed
to express the core concept, yet our definitions and methods apply within
communities, between communities, and for specific nodes.
This simplicity is directly linked to the generality of the definition.

To make our formulation concrete, we first discussed an existing edge
density community definition, the
absolute Potts model.  Methods based on Potts models have
historically proven to be very
accurate, but have shown limitations for weighted graphs.  To work
around this limitation, we proposed a new edge density model, the
variable topology Potts model, and shown its equivalence to the
absolute Potts model for unweighted graphs.  This generalization
points the way for community detection in all major types of graphs.
It is worth emphasizing that our edge density is
a \emph{local} community
definition, where nodes and communities are only affected by their
nearest neighbors.  This means our algorithm can be easily scaled to
large data sets, where only a small portion of the graph is
discoverable.

The core of this work involved developing criteria which must be
satisfied for any community assignment to be optimal.  The criteria are developed with respect to various changes in community
structure (community merging, addition of a node to a community, etc.)
being energetically favorable.  This is an important part of the evaluation of any
community definition or algorithm since it provides intuition as to
how the algorithms apply to real world graphs.  Furthermore, by developing
a set of criteria for various community characteristics, one can
 check that no unrealistic
properties emerge.  A chief example of such an unrealistic property, and
an example of the application of this technique, was Fortunato and
Barth\'elemy's investigation of a community merge criteria within the
modularity community definition in order to find that there was a
resolution limit\cite{fortunato2007resolution}.

We have applied our edge density methods to the AGM,
a recently proposed benchmark graph model\cite{yang2012structure}.  The creators of
this benchmark model claim that no currently existing method can solve
it, though it is likely that there exist methods other than edge density which also can.  We have solved it exactly, and demonstrated why we are able to do
so.
Our analysis of the AGM hints at the underlying reason for the
accuracy of the edge density community definition.
We postulate that the reason the edge density community definition
performs so well on
benchmark graphs is related to the fact that many benchmark graphs are
created with an
edge probability $p$ as the independent variable, to which our $\rho^*$ is
analogous.
Nevertheless, edge density methods have proven to be valuable when
applied to real world
networks\cite{ronhovde2011detecting,hu2011replica} and the latest
standard benchmark graphs (LFR, AGM)\cite{lancichinetti2009benchmarks,
  lancichinetti2009community,yang2012structure}.  This provides
evidence that edge densities, and the properties derived subsequently,
\emph{do} in fact reflect characteristics of important real world
networks.

\begin{acknowledgements}
  We would like to thank Santo Fortunato for careful reading and
  comments on this manuscript.  We would like to thank Dandan Hu for
  useful discussions.  RKD would like to thank the John and Fannie
  Hertz Foundation for research support via a Hertz Foundation
  Graduate Fellowship.  The work at Washington University in St. Louis
  has been supported by the National Science Foundation under NSF
  Grant DMR-1106293.  ZN also thanks the Aspen Center for Physics and
  the NSF Grant \#1066293 for hospitality.
\end{acknowledgements}

\appendix

\section{Simple edge density community detection algorithms}
\label{sec:ED-app-algorithms}

In the past few sections, we have outlined the ingredients necessary
for a community definition based on edge density.  We will now outline
several simple procedures for applying this definition to an
algorithm.  These dynamical procedures follow the precedent set by
existing literature, and our ability to use our definition, coupled
with multiple forms of dynamics, illustrates the separation between
community definitions and dynamics of community partitioning.

\subsection{Global method of Ronhovde, Hu, and Nussinov (RHN)}

This method takes an entire graph, and partitions it together,
resulting in a single community assignment for every node
\cite{ronhovde2009multiresolution, ronhovde2010local, hu2011phase,
hu2011replica}.  RHN use a
global $\rho^*$ (in the form of absolute Potts model $\gamma$) and
demonstrate an adaption to overlapping nodes.

In the RHN approach, we assign every node to a unique community consisting
of a single node as there are the same number of initial communities
as there are nodes.  Then, we make repeated passes of the following
changes in community assignments until we reach a point of local
stability: one in which none of the following moves will lower energy
any further.  More explicitly, the moves are:
\begin{itemize}
\item \emph{Local shifts.}  Choose one node, and change the community
  assignment of the node to another already-existing community.
  Accept if the change lowers the energy.  If the previous community
  only consisted of one node, that community vanishes and our number
  of communities $q$ shrinks by one.
\item \emph{New communities.}  Choose one node in a community of more
  than one node, and attempt to move it into a completely new
  community (which will then have only one node in it).  Accept if the
  change lowers the energy.  This increases the number of existing
  communities by one.
\item \emph{Merges.}  Attempt to merge two existing communities.
  Accept if the change lowers the energy.  This move contracts the
  number of communities by one.
\end{itemize}
The entire program outline above is a \emph{steepest-descent}
algorithm.  In order to get over
energy barriers, we perform $t$ independent trials with different
random seeds (for either in initial community assignments, or orders
of traversing nodes and communities in trials) as a means of gaining improved
sampling.  Typically, on the order of 5 trials are needed, and this is
found to be much more efficient of computer resources than simulated
annealing-like algorithms\cite{ronhovde2009multiresolution,
  ronhovde2010local, hu2011phase, hu2011replica}.

RHN also created an extension of this approach to overlapping communities\cite{ronhovde2011detecting}.
After performing the above steepest descent non-overlapping algorithm,
they take the following until they achieve a locally stable
configuration:
\begin{itemize}
\item \emph{Overlap expansion.}  For each community, attempt to add
  each node not currently in that community.  Do not remove the node
  from the previous community.  The number of communities stays
  constant, but one community gains an extra node.  Accept any
  additions which lower energy.
\item \emph{Overlap contraction.}  For each community and for each node
  in that community which was \emph{not} among the original nodes
  pre-expansion, attempt to remove that node from the community.
  Accept any removals which lower energy.
\end{itemize}

This algorithm has been shown to have exceptionally good performance
and efficiency.  A disadvantage of the original method is that it uses a global
$\rho^*$; more recently Ronhovde and Nussinov extended this approach to
inferring local structure \cite{ronhovde2012local}.

\subsection{Local algorithm of Lancichinetti, Fortunato, and Kert\'esz
  (LFK)}

In 2008, LFK developed a local fitness function and corresponding
dynamics for local community detection algorithms
\cite{lancichinetti2009detecting, lancichinetti2011finding, lee2010detecting}.  This function
shares some similarity with edge density, although is is distinct in using
external links as part of the measure of local fitness.  Their
dynamics can be easily adopted to our Potts models for edge density.

For this method, we choose a starting node from which to base our
community designations.  We then loop the following steps:
\begin{itemize}
\item For each community, take the set of all nodes adjacent to, but
  not within, the community.  Calculate the change in community fitness
  if each node were to be added to the community.  Add only the node
  which increases the fitness function by the most.  Repeat until no
  further additions can increase the fitness.
\item For each community, calculate the change in fitness if each node
  individually were to be removed.  If any node would increase fitness
  by its removal, remove the one node which most increases fitness.
  Repeat until no further nodes can increase fitness by their removal.
\item Once a locally stable community is found, repeat the procedure
  starting from another node which has not yet been assigned to any
  community.
\end{itemize}

This procedure allows overlapping communities to be found, and allows
a locally tunable $\rho^*$, as opposed to a global $\rho^*$.  This
method has not been applied to our edge density community definition,
but can easily be via local optimization of
Eq.~(\ref{eq:APMHamiltonian}).  This algorithm provides a method of
community detection when only a small portion of the graph is
visible.

\subsection{Advanced methods}

Methods such as simulated annealing or heat bath algorithms are
extensions to the above methods \cite{reichardt2004detecting,
  hu2011phase}.  As the edge density community definition is based on
a Hamiltonian and not a set of dynamic steps, we
have the freedom to choose any dynamical steps we may like to minimize
energies.  Energy optimization has been extensively studied in the physics
literature, and there are many lessons which can be taken from spin
glasses, molecular dynamics, and other fields.

It would be useful to study the ability of various methods to overcome
local energy barriers.  For example, RHN noticed that community merge
moves were important in order to surmount local
barriers\cite{nussinovunpublished}.  Without
these, singe-node community shifts were not able to effectively lower
energies.  There are subtle differences between the order of
additions and removals of the RHN and the LFK methods, which could
have impact on the performances of the minimizations.  These methods
will not be discussed further here.

\subsection{Multi-replica inference}
\label{sec:ED-multireplica}

In order to use the methods outlined above, we must know $\rho^*$
before we
begin our detection process.  Since this, in general, can not be known
in advance, we need to infer $\rho^*$ via some technique.  There is an
established procedure for this \emph{multi-resolution analysis} in the
literature\cite{ronhovde2010local}.

To do this, we \emph{scan} across a range of values and perform
multiple community detections (``$r$ replicas'') at each $\rho^*$.  We
infer that if the results from community detections at a given
$\rho^*$ are very similar, we have good community detection.  This is
equivalent to saying that at good values of $\rho^*$, we have one
dominant community assignment that is uniformly detected.  Past work
has shown this to be a very good procedure
\cite{ronhovde2009multiresolution, ronhovde2010local, arenas2008analysis}.

There are various measures of inter-replica similarity, most based on
the concepts of information
theory\cite{traud2011comparing,mackay2003information,
  rosvall2007information}.  Various
proposed choices include the variance of information ($VI$)
\cite{karrer2008robustness, ronhovde2009multiresolution,meilua2007comparing},
normalized mutual information
($I_N$)\cite{ronhovde2009multiresolution,danon2005comparing}, and a
generalized normalized
mutual information capable of handling overlaps ($N$)
\cite{lancichinetti2009detecting}.  In this work,
we will use a version of the F-score, $F_1$ generalized to handle
partitions (see Appendix \ref{sec:thesis-F1})
\cite{lee2010detecting,hu2011replica}.  The unifying characteristic
of these measures is that they take two complete community assignments
and output a number which indicates the similarity of the partitions.
Most measures, including the F-score, are normalized as follows: a value
of unity indicates perfect agreement in partitions, while a value of
zero indicates completely decorrelated community assignments.

Our F-score development extends the concept
of comparing two partitions of the system to overlapping nodes,
similar to, but conceptually simpler than, the developments of
Lancichinetti \textit{et. al.}\cite{lancichinetti2009detecting}.  Our
derivation provides more insight into the actual performance of the
community detection algorithm.  Full details are located in
Appendix~\ref{sec:thesis-F1}.  There, we derive a modification of
$F_1$ for comparing entire partitions (instead of single communities)
to each other, which we denote $F_1^R$.  $F_1^R$ has the
following interpretation: a value of unity indicates perfect agreement
between community assignments, and a value approaching zero indicates
perfect decorrelation of community structure.

These similarity measures ($VI$, $I_N$, $F_1^R$, etc.) are also used
for testing the outcomes of community detection experiments.  If we
know the correct community assignments (community assignment $R_0$),
we expect the similarity between the detected and correct communities
to be unity when averaged across the detected configurations $R_1,
R_2, \ldots , R_r$:
\begin{equation}
  S^0 = \frac{1}{r} \sum_{i=1}^r S(R_0, R_i)
\end{equation}
using similarity measure $S$.
To compute the inter-replica symmetry, we
use
\begin{equation}
  S^{IR} = \frac{1}{r(r-1)} \sum_{i=1}^r \sum_{j\neq i}^r  S(R_i, R_j).
\end{equation}

Thus we can state our general criteria for a useful multi-resolution
algorithm: as a function of $\rho^*$, we must have a maximum of
$S^{IR}$ at the same locations as $S^0$ unity, in order to have both
\emph{accurate community detection} and \emph{the ability to infer
  correct $\rho^*$ with no prior information}.

\section{Why do current methods not properly handle dense overlaps?}

One of the claims of YL is that current community detection methods do
not properly handle dense overlaps as in
Fig.~\ref{fig:edgedensityoverlap} \cite{yang2012structure}.  In this
section, we will explain why that is for certain popular methods.  For
our thought experiment, consider two communities $A$ and $B$, with a
dense overlap as shown in that figure.

\subsection{Clique percolation}

In clique percolation, a $n$-clique (set of $n$ nodes all mutually
connected) is located within a graph\cite{palla2005uncovering}.  Then, all cliques which overlap
$n-1$ nodes of the first clique are identified.  This process
continues, and a community is defined as set of all nodes reachable by
this percolation process.  For both communities $A$ and $B$, the
percolating clique will enter the overlap and detect these nodes.
Once that happens, the percolating clique can ``jump'' to the other
community, and they will merge together.

\subsection{Betweenness}

When using betweenness to detect communities, shortest paths (along
graph edges) are drawn between \emph{all} pairs of
nodes\cite{newman2004finding}.  The edges with the most shortest paths
falling through them are considered the boundary between communities,
and are virtually removed, in sequence, until isolated communities
remain.  An external criteria (such as modularity) is used to
determine the stopping point of this process.  In the dense overlap
viewpoint, there is no region with fewer edges, which means that edges
at a community boundary never have many paths focused through them.
Betweenness is thus unable to determine community boundaries.

\subsection{Other algorithms}

Thought experiments similar to the above can be performed for other
community detection algorithms.  While YL claim that existing
community detection algorithms can not handle dense overlaps, certain
methods can do so.  As already explained above, the Absolute Potts
Model is able to detect sparse
overlaps\cite{ronhovde2009multiresolution}.  Furthermore, other
various local optimization algorithms are able to detect communities
in the face of dense overlaps\cite{lancichinetti2009detecting,
  lancichinetti2011finding}.

\section{F-score partition similarity metric}
\label{sec:thesis-F1}
\input{F1.tex}

\bibliographystyle{apsrev}
\bibliography{/home/richard/rkd/doc/bib/cmty.bib,%
  /home/richard/rkd/doc/bib/cmty-thesis.bib}{}

\end{document}

%% file: F1.tex
In order to be able to quantify the performance of community detection
algorithms, we must have tools to compare the similarity two
partitions (or more generally, \emph{covers}) which allow
overlaps).  Furthermore, one of the core tenants of our multi-resolution
algorithm is that high uniformity across replicas indicates good
community detection solutions.  There are a variety of these functions
derived from information theory which answer the question ``if you
know one partition of the system, what is known about replicas?''.

This F-score development discussed here
has certain advantages.  First, it can handle
overlapping communities.  It also can handle incomplete partitions, where
the communities being detected or sought consist of only a portion of
the nodes of the entire graph, as opposed to other functions which
only compare complete partitions of the graph.  Thus, the F-score is a
valuable tool for local community detection study.  The F-score has the
interpretation: $F=1$ implies that we have exactly recovered
a known community, with every node detected and no extraneous nodes
detected.

\subsection{Single-community F-score}

\begin{figure}
  \centering
  (a) $|A|=12$, $|A'|=10$\\
  \includegraphics[width=\threehighwidth]{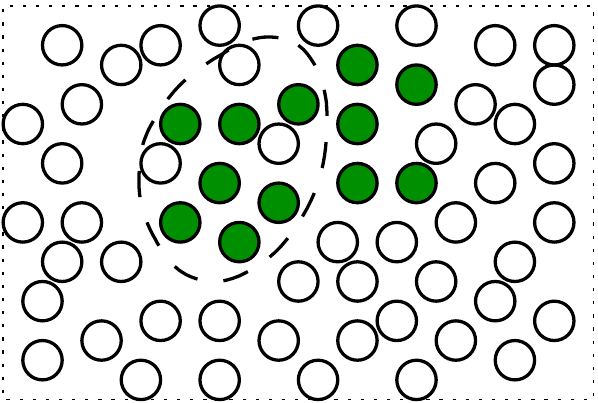}

  (b) precision = 7/10\\
  \includegraphics[width=\threehighwidth]{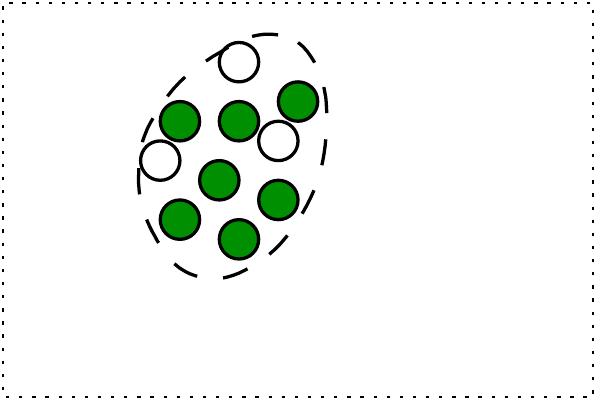}

  (c) recall = 7/12\\
  \includegraphics[width=\threehighwidth]{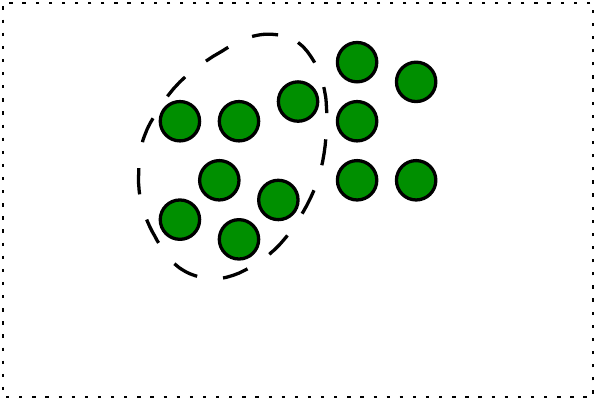}
  \caption[F-score calculation]{Example for calculation of F-score.
    (a) We have a known community $A$ (green nodes), and the results
    from the community detection algorithm $A'$ overlaid (dashed
    oval).  (b) Calculation of precision: 7/10 detected nodes are
    correct.  (c) Calculation of recall: 7/12 correct nodes are
    detected.}
  \label{fig:fscore-demo}
\end{figure}

As an example, let us consider the situation depicted in
Fig.~\ref{fig:fscore-demo}.  We have a community $A$ which we want to
detect.  We apply some algorithm and end up with a group of nodes
$A'$ (dark and green).  We see there are 10 nodes in $A$, $|A|=10$.  Our
algorithm has returned 12 nodes, $|A'|=12$.  Note that we have an
overlap of seven nodes, which means five nodes were detected which are
incorrect, and we missed three nodes we should have detected.

We will consider our initial development to be
for a single community (in other contexts, F-score is defined only for
single-group searching).  Using some local community detection
algorithm, we detect a group of nodes.  We use the phrase
``community'' to indicate the \emph{known} group of nodes we want to
match.  We use the term ``detected nodes'' to indicate what the
community detection algorithm actually returns.  Note that there are
few published local community detection algorithms of this type (that
will return only a single community in isolation, as opposed to a
partition/cover of the entire system).  This is not a practical limitation,
as we develop methods of averaging to compensate for this.

There are two components to the F-score.  First is the
\emph{precision}, measuring how many nodes in the test community
actually belong in the known community.  It answers the question ``of all nodes
detected, how many of them are relevant to understanding the
characteristic of the community''.  Precision is defined
\begin{equation}
  \label{eq:FscorePrecision}
  \text{precision}(A, A')
  = \frac{\text{correct nodes returned}}{\text{total nodes returned}}
  = \frac{|A \cap A'|}{|A'|}
\end{equation}
The nomenclature $|A|$ indicates the number of nodes in $A$, and
$\cap$ indicates set intersection.  In Fig.~\ref{fig:fscore-demo},
precision$=7/10$.  A precision of 1 indicates that every node detected
is in the community.  A precision of zero means that no detected nodes
are in the community.  A lower precision indicates more false
positives.

Next, the \emph{recall} indicates what fraction of the community nodes
were actually detected.  It answers the question ``of all nodes in the
community
(that we want to detect), what fraction were detected?''  Recall is defined as
\begin{equation}
  \label{eq:FscoreRecall}
  \text{recall}(A, A')
  = \frac{\text{correct nodes returned}}{\text{total community nodes}}
  = \frac{|A \cap A'|}{|A|}
\end{equation}
In Fig.~\ref{fig:fscore-demo}, recall$=7/12$.  A recall of 1 indicates
that we have managed to not miss any nodes in the desired community (though
there could be excess nodes returned, too).  A recall of less than one
indicates false negatives.

Generally, there is a trade-off between precision and recall.
Precision can be made 1 by selecting fewer nodes (in the limiting
case, by selecting only one node which is known to be in the
community), at the cost of a very low recall.  Conversely, recall can
be made 1 by selecting every node in the graph, at the expense of a
low precision.  Overall goodness of our search process is measured in
the form of the \emph{F-score}, the weighted harmonic mean of
precision and recall
\begin{equation}
  \label{eq:Fscore}
  F_\beta(A, A') = (1+\beta^2)
  \frac{\text{prec}(A, A')\text{recall}(A, A')}{\beta^2\text{prec}(A,
    A')+\text{recall}(A, A')}.
\end{equation}
The parameter $\beta$ weights the relative importance of precision and recall, with a higher
$\beta$ weighting recall more.  This selectivity has a benefit in
community detection work: if the user has a preference for ensuring
detection of all nodes at a possible cost of false positives, or the
converse, that can be accommodated.
\begin{figure}
  \centering
  \includegraphics[width=\wholewidth]{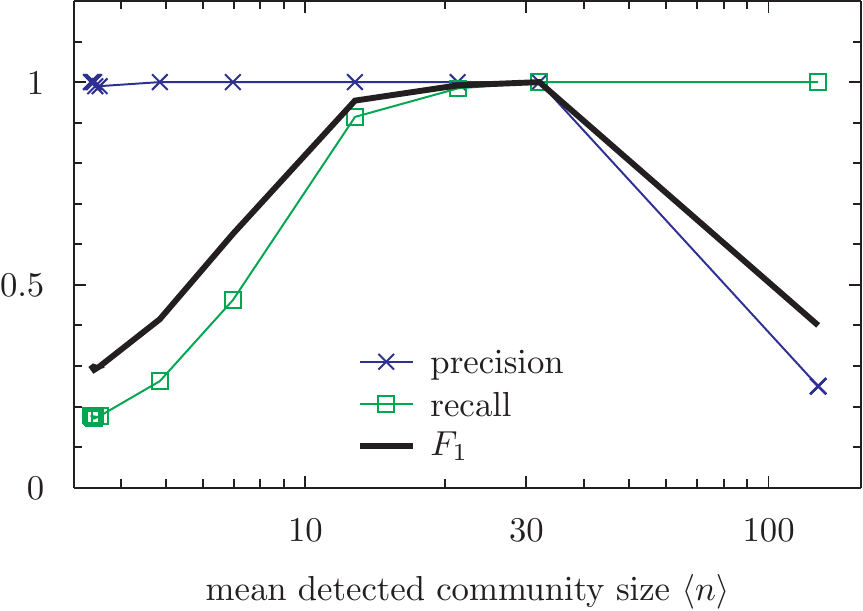}
  \caption[Precision vs recall]{
    Trade-off between precision and recall.  At low mean
    community size $n$, our detected communities are too small and we
    have high precision but low recall.  At larger community size, our
    detected communities are too large and we have low precision but
    high recall.  When we properly detect the correct communities,
    both precision and recall are high, and we have $F_1=1$.  In this
    sample graph, actual communities have a size of $n=32$ which we
    can observe from the peak of $F_1$.  This figure actually plots
    $F_1^0$ which will be explained in the next sections, but
    conceptually the results are the same.  Tested on $q=4$ $n=32$ SBM
    graph with $p_\mathrm{in}=.5$ and $p_\mathrm{out}=.1$.}
  \label{fig:fscore-precrecall-tradeoff}
\end{figure}

In our work, we will use $F_1$, the balanced score.  $F_1$ is the
harmonic mean of precision and recall,
\begin{equation}
  \label{eq:F1}
  F_1(A, A') =
  2 \frac{\text{prec}(A, A')\text{recall}(A, A')}{\text{prec}(A,
    A')+\text{recall}(A, A')}.
\end{equation}
The use of $F_1$ is between a set of nodes (the known ``community''
$A$) and another set (the ``detected nodes'', $A'$), and for
$\beta\neq1$ is non-symmetric in $A$ and $A'$.  For $\beta=1$, $F_1$
is symmetric, with precision and recall swapping values when $A$ and
$A'$ are swapped.

The overall F-score in Fig.~\ref{fig:fscore-demo} is
$\frac{2\cdot\frac{7^2}{10\cdot12}}{\frac{7}{10}+\frac{7}{12}} \approx
.64$.  Note that the F-score does \emph{not} depend on the total
number of nodes in the graph.  This means our F-score scale does not
change as we increase the total graph size (at approximately constant
community size), providing theoretical advantages for \emph{local} community
detection.

\subsection{Partition F-score}

In order to compare two partitions, we take one partition as the
``known'' community assignment we want to match ($\alpha$) and another
partition as the result from our (now global) community detection
algorithm.  We take an average of $F_1$ scores,
\begin{equation}
  \label{eq:FscorePartitions}
  F_1^P(\alpha, \alpha') =
  \frac{1}{|\alpha|} \sum_{A \in \alpha}
  \max_{A' \in \alpha'}\left( F_1(A, A') \right)
\end{equation}
In words, for every community $A$ in the known partition, we compute
the F-score with all communities $A'$ in $\alpha'$ and, and choose
the one which maximizes the $F_1$ with $A$.  We take the average value
of all of these maxima.  The number of communities in $\alpha$ is
represented by $|\alpha|$.

As defined above, $F_1^P$ is non-symmetric.  In the final value, each
community in $\alpha$ is used at least once as the first argument to
$F_1$, but the only communities in $\alpha'$ used as the second
argument are those which maximize $F_1$ to at least one of the
communities in $\alpha$.  There can be communities in $\alpha'$ which
are left unused and do not affect the result.

It is extremely important to understand the non-symmetric nature of
$F_1^P$.  As it is posed above, $F_1^P$ is a useful metric if every
known community is matched by at least one detected community.  It
answers the question ``Is every community detected at least once?''.
A distinct question is ``does every community in $\alpha'$
correspond to at least one real community?''.  To answer this second
question, we swap the roles of $\alpha$ and $\alpha'$ in $F_1^P$ and
instead compute $F_1^P(\alpha', \alpha)$.  Just like precision and
recall both have their uses in understanding the community detection
process, $F_1^P(\alpha', \alpha)$ and $F_1^P(\alpha, \alpha')$ both
tell different and useful properties of our minimization: One answers
the question ``is every community detected at least once?'', the other
answers ``does every detected community represent a real community?''

Naively, $F_1^P$ is an $O(|\alpha| |\alpha'|)$ calculation, because we
must compare every community in $\alpha$ to every community in
$\alpha'$.  There is potential for optimization by first generating a
list of only overlapping communities.  Each actual $F_1$ evaluation
consists only of the operations of set intersection and set
cardinality, which can be made efficient.

Analogously, we could define partition precision/recall as
\begin{eqnarray}
  \text{prec}^P(\alpha, \alpha') &=&
  \frac{1}{|\alpha|} \sum_{A \in \alpha}
  \max_{A' \in \alpha'}\left( \text{prec}(A, A') \right) \\
  \text{recall}^P(\alpha, \alpha') &=&
  \frac{1}{|\alpha|} \sum_{A \in \alpha}
  \max_{A' \in \alpha'}\left( \text{recall}(A, A') \right).
\end{eqnarray}
A low precision generally indicates that communities are being
detected too large or too liberally.  A low recall generally indicates
that communities are being detected too small or too conservatively.
This can provide valuable information to monitor the performance of
our algorithm.  Note that $F_1^{P}$ is not communative with respect
to the averaging and the precision/recall,
\begin{equation}
  F_1^{P} \neq 2\cdot \frac{\text{prec}^{P}\text{recall}^{P}}{\text{prec}^{P}+\text{recall}^{P}}
\end{equation}

\subsection{F-score applications}

We apply $F_1^P$ in two ways: first to, compare the uniformity of a
handful of replicas (results from community detection applied to the
same graph with different initial conditions/random seeds).
Second, to compare partitions with a known state.  For these, we assume
that we have $r$ replicas $\alpha_1, \alpha_2, \ldots$.

First, we can use $F_1^P$ as a measure of inter-replica uniformity in
our multi-resolution algorithm.  This takes the place of the variance
of information, normalized mutual information, or $N$-measure.  For
this, we calculate $F_1^P$ with respect to every pair of replicas,
\begin{equation}
  F_1^{IR}(\alpha,\beta)
  = \frac{1}{r-1} \sum_{\alpha\neq\beta}F_1^R(\alpha, \beta),
\end{equation}
for the partitions $\alpha$ and $\beta$ in the replicas.  This
measure is 1 when all replicas are identical.  Note that we sum over
both the argument orders $(\alpha, \beta)$ and $(\beta, \alpha)$ due
to the non-symmetric nature of the $F_1^P$ measure.

\begin{figure}
  \centering
  \includegraphics[width=\wholewidth]{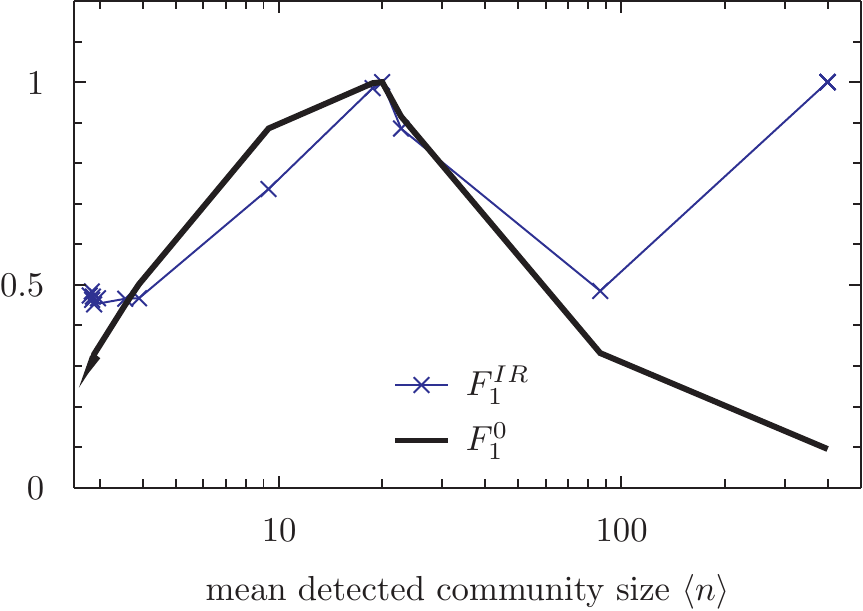}
  \caption[Validation of $F_1^{IR}$ and $F_1^{0}$ correlation for
  multi-resolution algorithms]{
    Comparison between $F_1^{IR}$ and $F_1^0$ on a $q=20$
    $n=20$ SBM graph.  We see that the measures initially reach a
    maximum at the
    same point $n=20$, allowing us to use $F_1^{IR}$ to infer that
    communities have a size of 20 nodes.  The $n=400$ maximum of
    $F_1^{IR}$ is the trivial result that when one giant community is
    detected, the result is very uniform.}
  \label{fig:fscore-IRvs0}
\end{figure}

\begin{figure}
  \centering
  \includegraphics[width=\wholewidth]{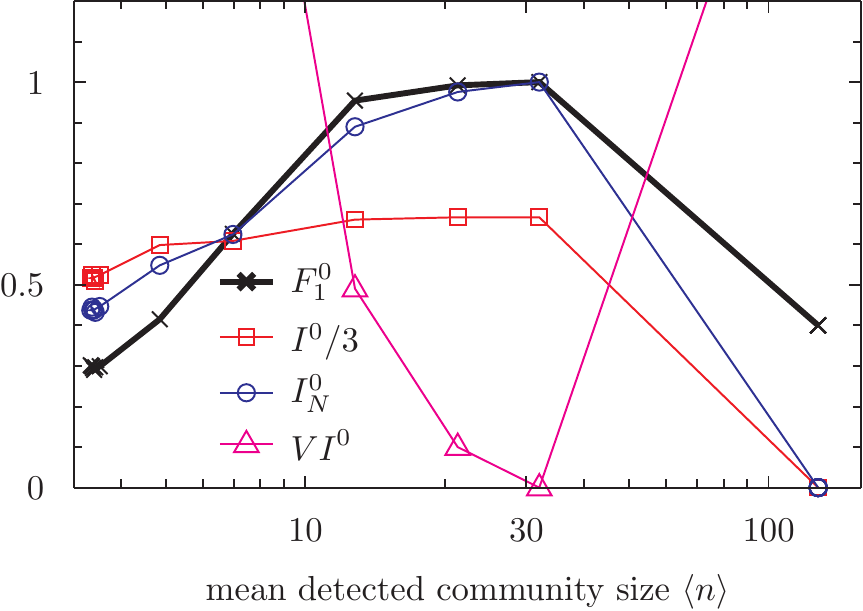}
  \caption[Comparison of F-score to other similar measures]{
    Comparison of $F_1^0$ and other
    partition measures on the same $q=4$ $n=32$ SBM graph.  We see
    that all measures contain extrema at the same community size
    $\left<n\right>$, validating that all measures convey
    approximately the same information on this graph.  We compare
    mutual information
    $I^0$\cite{ronhovde2009multiresolution,danon2005comparing},
    normalized mutual information $I_N$
    \cite{ronhovde2009multiresolution,danon2005comparing}, and
    variance of information $VI$\cite{ronhovde2009multiresolution,karrer2008robustness,meilua2007comparing}. }
  \label{fig:fscore-compare}
\end{figure}

Next, we can form a version of this measure which detects similarity
to a known structure which we designate as $\alpha_0$.  The $F_1^P$
score with respect to a known configuration is defined as
\begin{equation}
  F_1^0 = \frac{1}{r} \sum_\alpha F_1^P(\alpha_0, \alpha)
\end{equation}
for all replicas $\alpha$.  Due to the asymmetric nature of $F_1^P$,
this measure indicates the extent to which all known communities are
detected in replicas, but not the extent to which all detected
communities represent real structure.  The asymmetric nature of $F_1$
here is beneficial, because we can handle cases where we partition the
entire graph into communities, but can also search for detection of
only a few partitions within a larger graph.
In Fig.~\ref{fig:fscore-IRvs0}, we show that $F_1^{IR}$ is useful for inferring
$F_1^0$, and in Fig.~\ref{fig:fscore-compare}, we show that $F_1^0$ behaves
similarly to other partition comparison functions.

\subsection{Discussion}

The F-score is a tool for comparing partitions sharing some similarity
with already existing measures, but with the possibility for further
extension to local community detection and overlapping nodes.  By
combining different orders applications of averaging, precisions,
recalls, and F-scores, it provides additional insight to the internal
workings of CD methods. 